\documentclass{lmcs}
\pdfoutput=1

\usepackage{lastpage}
\lmcsdoi{18}{2}{1}
\lmcsheading{}{\pageref{LastPage}}{}{}%
{Apr.~30,~2021}{Apr.~13,~2022}{}


\usepackage{unixode}
\usepackage[T1]{fontenc}

\usepackage{microtype}

\usepackage{amssymb,amsmath}
\newcommand\citet[1]{\cite{#1}}
\usepackage[misc]{ifsym} 

\usepackage{cite}

\usepackage{color}
\definecolor{keywordcolor}{rgb}{0.5, 0.0, 0.0}   
\definecolor{commentcolor}{rgb}{0.4, 0.4, 0.4}   
\definecolor{symbolcolor}{rgb}{0.0, 0.1, 0.6}    
\definecolor{sortcolor}{rgb}{0.0, 0.4, 0.0}      

\usepackage{listings,upgreek,upquote}

\usepackage{hyperref}
\usepackage[capitalise]{cleveref} 

\begin{document}
\title[Beyond Notations]{Beyond Notations: Hygienic Macro Expansion\texorpdfstring{\\}{ }for Theorem Proving Languages\rsuper*}
\titlecomment{{\lsuper*}This is an extended version of \citet{ullrich2020beyond}, with the most significant changes being the addition of \cref{sec:quot} and \cref{sec:eager}.}
\author[S.~Ullrich]{Sebastian Ullrich}
\address{Karlsruhe Institute of Technology, Germany}
\email{\texttt{sebastian.ullrich@kit.edu}}

\author[L.~de Moura]{Leonardo de Moura}
\address{Microsoft Research, USA}
\email{\texttt{leonardo@microsoft.com}}

\begin{abstract}
  In interactive theorem provers (ITPs), extensible syntax is not only crucial to lower the cognitive burden of
  manipulating complex mathematical objects, but plays a critical role in developing reusable abstractions in libraries.
  Most ITPs support such extensions in the form of restrictive ``syntax sugar'' substitutions and other ad hoc mechanisms,
  which are too rudimentary to support many desirable abstractions. As a result, libraries are littered with unnecessary
  redundancy. Tactic languages in these systems are plagued by a seemingly unrelated issue: accidental name capture,
  which often produces unexpected and counterintuitive behavior. We take ideas from the Scheme family of programming
  languages and solve these two problems simultaneously by proposing a novel \emph{hygienic macro system} custom-built for ITPs. We further
  describe how our approach can be extended to cover type-directed macro expansion resulting in a single, uniform system
  offering multiple abstraction levels that range from supporting simplest syntax sugars to elaboration of formerly
  baked-in syntax. We have implemented our new macro system and integrated it into the new version of the Lean
  theorem prover, Lean 4. Despite its expressivity, the macro system is simple enough that it can easily be integrated into
  other systems.
\end{abstract}

\maketitle              

\section{Introduction}
\label{sec:intro}

\emph{Mixfix} notation systems have become an established part of many modern ITPs for attaching terse and familiar syntax
to functions and predicates of arbitrary arity.

\begin{minipage}{0.8\linewidth}
\begin{lstlisting}
_⊢_:_ = Typing
Notation "Ctx ⊢ E : T" := (Typing Ctx E T).
notation typing ("_ ⊢ _ : _")
notation Γ `⊢` e `:` τ := Typing Γ e τ
\end{lstlisting}
\end{minipage}
\begin{minipage}{0.1\linewidth}
  \begin{flushright}
    \small\itshape
    Agda \\
    Coq \\
    Isabelle \\
    Lean 3
  \end{flushright}
\end{minipage}

As a further extension, all shown systems also allow \emph{binding names} inside mixfix notations.

\begin{minipage}{0.8\linewidth}
\begin{lstlisting}
syntax ∃ A (λ x → P) = ∃[ x ∈ A ] P
Notation "∃ x , P" := (exists (fun x => P)).
notation exists (binder "∃")
notation `∃` binder `,` r:(scoped P, Exists P) := r
\end{lstlisting}
\end{minipage}
\begin{minipage}{0.1\linewidth}
  \begin{flushright}
    \small\itshape
    Agda \\
    Coq \\
    Isabelle \\
    Lean 3
  \end{flushright}
\end{minipage}

While these extensions differ in the exact syntax used, what is true about all of them is that at the time
of the notation declaration, the system already, statically knows what parts of the term are bound by the newly
introduced variable. This is in stark contrast to \emph{macro systems} in Lisp and related languages where the
expansion of a \emph{macro} (a syntactic substitution) can be specified not only by a \emph{template expression} with placeholders like above, but also by
arbitrary \emph{syntax transformers}, i.e.\ code evaluated at compile time that takes and returns a syntax
tree.\footnote{These two macro declaration styles are commonly referred to as \emph{pattern-based} vs.\ \emph{procedural}.} As we
move to more and more expressive notations and ideally remove the boundary between built-in and user-defined
syntax, we argue that we should no more be limited by the static nature of
existing notation systems and should instead introduce syntax transformers to the world of ITPs.

However, as usual, with greater power comes greater responsibility. By using arbitrary syntax transformers, we lose the
ability to statically determine what parts of the macro template can be bound by the macro input (and vice versa). Thus
it is no longer straightforward to avoid \emph{hygiene} issues (i.e.\ accidental capturing of identifiers; \cite{kohlbecker1986hygienic}) by automatically
renaming identifiers. We propose to learn from and adapt the macro hygiene systems
implemented in the Scheme family of languages for interactive theorem provers in order to obtain more general but still well-behaved
notation systems.

After giving a practical overview of the new, macro-based notation system we
implemented in Lean 4~\cite{demoura2021lean} in \cref{sec:new}, we describe the
issue of hygiene and our general hygiene algorithm, which should be just as applicable to other ITPs, in
\cref{sec:hygge}. \cref{sec:impl} gives a detailed description of the implementation of this algorithm, and macros in general, in Lean 4. In
\cref{sec:elab}, we extend the use case of macros from mere syntax substitutions to type-aware elaboration.%
\footnote{By ``elaboration'', we mean the transformation of surface-level syntax into the explicit kernel term/declaration representation, including type and typeclass inference.}
Finally, we have already encountered hygiene issues in the current version of Lean in a different part of the system:
the tactic framework. We discuss how these issues are inevitable when implementing reusable tactic scripts and how our
macro system can be applied to this hygiene problem as well in \cref{sec:tactic}.

\emph{Contributions}. We present a system for hygienic macros optimized for theorem proving languages as implemented\footnote{\url{https://github.com/leanprover/lean4/blob/IJCAR20-LMCS/src/Lean/Elab}} in
the new version of the Lean theorem prover, Lean 4.
\begin{itemize}
\item We describe a novel, efficient hygiene algorithm 
  to employ macros in ITP languages at large: a combination
  of a white-box, effect-based approach for detecting newly introduced identifiers and an efficient encoding of scope
  metadata.
\item We show how such a macro system can be seamlessly integrated into existing elaboration designs to support
  type-directed expansion even if they are not based on homogeneous source-to-source transformations.
\item We show how hygiene issues also manifest in tactic languages and how they can be solved with the same
  macro system. To the best of our knowledge, the tactic language in Lean 4 is the first tactic language in an
  established theorem prover that is automatically hygienic in this regard.
\end{itemize}

\section{The New Macro System}
\label{sec:new}

Lean's previous notation system as shown in \cref{sec:intro} is still supported in Lean 4, but based on a much more general
macro system; in fact, the \lstinline{notation} command\footnote{A Lean file is made up of a sequence of commands, which are processed in turn and can extend the environment with new declarations or metadata (such as custom parsers).} itself has been reimplemented as a macro, more specifically as a
\emph{macro-generating macro} making use of our tower of abstraction levels. The corresponding Lean 4
command\footnote{All examples including full context can be found in the supplemental material at
  \url{https://github.com/Kha/macro-supplement}.}
for the example from the previous section
\begin{lstlisting}[linerange=typing-end]
import Lean  -- necessary only for the last tactic example

section «2»
namespace Typing
axiom Ctxt : Type
axiom Term : Type
axiom Typ  : Type
axiom Typing : Ctxt → Term → Typ → Type
-- typing
notation Γ "⊢" e ":" τ => Typing Γ e τ
-- end
#check fun a b c => a ⊢ b : c
--macro Γ:term "⊢" e:term ":" τ:term : term => `(Typing $Γ $e $τ)
--
--syntax term "⊢" term ":" term : term
--macro_rules
--  | `($Γ ⊢ $e : $τ) => `(Typing $Γ $e $τ)
end Typing

-- already built into Lean
/-
-- exists
notation "∃" b "," P => Exists (fun b => P)
-- end
-/
#check ∃ x, x = x
#check ∃ (x : Nat), x = x


-- defthunk
macro "defthunk" id:ident ":=" e:term : command =>
  `(def $id := Thunk.mk (fun _ => $e))
defthunk big := mkArray 100000 true
-- end
#check big


namespace Union
abbrev Set (α : Type) := α → Prop
axiom setOf {α : Type} : (α → Prop) → Set α
axiom mem {α : Type} : α → Set α → Prop
axiom univ {α : Type} : Set α
axiom Union {α : Type} : Set (Set α) → Set α
macro:100 x:term " ∈ " s:term:99 : term => `(mem $x $s)

-- union
syntax "{" term "|" term "}" : term
macro_rules
  | `({$x ∈ $s | $p}) => `(setOf (fun $x => $x ∈ $s ∧ $p))
  | `({$b      | $p}) => `(setOf (fun $b => $p))

notation "⋃" b "," p => Union {b | p}
-- end

#check ⋃ x,              x = x
#check ⋃ (x : Set Unit), x = x
#check ⋃ x ∈ univ,       x = x


-- le
macro_rules
  | `({$x ≤ $e | $p}) => `(setOf (fun $x => $x ≤ $e ∧ $p))
-- end

#check {x ≤ 1 | x = x}
end Union


-- final example: see ./Bigop.lean
end «2»


section «3»
-- see also ./Expander.lean

-- hygiene_example
def x := 1
def e := fun (y : Nat) => x
notation "const" e => fun (x : Nat) => e
def y := const x
-- end
#check y


-- hygiene_example2
macro "m" n:ident : command => `(
  def f := 1
  macro "mm" : command => `(
    def $n := f + 1
    def f := $n + 1))
-- end
m f
mm
mm
end «3»


section «6»
-- myTac
macro "myTac" : tactic => `(intro h; exact h)
theorem triv (p : Prop) : p → p := by myTac
-- end

open Lean.Elab.Tactic
-- myTac2
def myTac2 : TacticM Unit := do
  let stx ← `(tactic| intro h; exact h)
  evalTactic stx
-- end
end «6»
\end{lstlisting}
expands to the macro declaration
\begin{lstlisting}
macro Γ:term "⊢" e:term ":" τ:term : term => `(Typing $Γ $e $τ)
\end{lstlisting}
where the \emph{syntactic category} (\lstinline{term}) of placeholders and of the entire macro is now specified explicitly.
The right-hand side uses an explicit \emph{syntax quasiquotation} to construct the syntax tree,
with syntax placeholders (\emph{antiquotations}) prefixed with \lstinline{$}. As suggested by the
explicit use of a quotation, the right-hand side may now be an arbitrary Lean term computing a syntax object; in other words,
there is no distinction between pattern-based and procedural macros in our system. We can now use this abstraction level to
implement simple macros in syntactic categories other than \lstinline{term}, such as for commands.
\begin{lstlisting}[linerange=defthunk-end]
import Lean  -- necessary only for the last tactic example

section «2»
namespace Typing
axiom Ctxt : Type
axiom Term : Type
axiom Typ  : Type
axiom Typing : Ctxt → Term → Typ → Type
-- typing
notation Γ "⊢" e ":" τ => Typing Γ e τ
-- end
#check fun a b c => a ⊢ b : c
--macro Γ:term "⊢" e:term ":" τ:term : term => `(Typing $Γ $e $τ)
--
--syntax term "⊢" term ":" term : term
--macro_rules
--  | `($Γ ⊢ $e : $τ) => `(Typing $Γ $e $τ)
end Typing

-- already built into Lean
/-
-- exists
notation "∃" b "," P => Exists (fun b => P)
-- end
-/
#check ∃ x, x = x
#check ∃ (x : Nat), x = x


-- defthunk
macro "defthunk" id:ident ":=" e:term : command =>
  `(def $id := Thunk.mk (fun _ => $e))
defthunk big := mkArray 100000 true
-- end
#check big


namespace Union
abbrev Set (α : Type) := α → Prop
axiom setOf {α : Type} : (α → Prop) → Set α
axiom mem {α : Type} : α → Set α → Prop
axiom univ {α : Type} : Set α
axiom Union {α : Type} : Set (Set α) → Set α
macro:100 x:term " ∈ " s:term:99 : term => `(mem $x $s)

-- union
syntax "{" term "|" term "}" : term
macro_rules
  | `({$x ∈ $s | $p}) => `(setOf (fun $x => $x ∈ $s ∧ $p))
  | `({$b      | $p}) => `(setOf (fun $b => $p))

notation "⋃" b "," p => Union {b | p}
-- end

#check ⋃ x,              x = x
#check ⋃ (x : Set Unit), x = x
#check ⋃ x ∈ univ,       x = x


-- le
macro_rules
  | `({$x ≤ $e | $p}) => `(setOf (fun $x => $x ≤ $e ∧ $p))
-- end

#check {x ≤ 1 | x = x}
end Union


-- final example: see ./Bigop.lean
end «2»


section «3»
-- see also ./Expander.lean

-- hygiene_example
def x := 1
def e := fun (y : Nat) => x
notation "const" e => fun (x : Nat) => e
def y := const x
-- end
#check y


-- hygiene_example2
macro "m" n:ident : command => `(
  def f := 1
  macro "mm" : command => `(
    def $n := f + 1
    def f := $n + 1))
-- end
m f
mm
mm
end «3»


section «6»
-- myTac
macro "myTac" : tactic => `(intro h; exact h)
theorem triv (p : Prop) : p → p := by myTac
-- end

open Lean.Elab.Tactic
-- myTac2
def myTac2 : TacticM Unit := do
  let stx ← `(tactic| intro h; exact h)
  evalTactic stx
-- end
end «6»
\end{lstlisting}
\lstinline{macro} itself is another command-level macro that, for our \lstinline{notation} example, expands to two commands
\begin{lstlisting}
syntax term "⊢" term ":" term : term
macro_rules
  | `($Γ ⊢ $e : $τ) => `(Typing $Γ $e $τ)
\end{lstlisting}
that is, a pair of parser extension (which we will not further discuss in this paper) and syntax transformer.
Our reason for ultimately separating these two concerns is that we can now obtain a well-structured syntax tree pre-expansion, i.e.\ a
\emph{concrete} syntax tree, and use it to implement source code tooling such as auto-completion, go-to-definition, and
refactorings. Implementing even just the most basic of these tools for the Lean 3 frontend that combined parsing and
notation expansion meant that they had to be implemented right inside the parser, which was not an extensible or even
maintainable approach in our experience.

Both \lstinline{syntax} and \lstinline{macro_rules} are in fact further macros for regular Lean definitions encoding
procedural metaprograms, though users should rarely need to make use of this lowest abstraction level explicitly. Both
commands can only be used at the top level; we are not currently planning support for local macros.

There is no more need for the complicated \lstinline{scoped} syntax since the desired translation can now be specified
naturally, without any need for further annotations.
\begin{lstlisting}[linerange=exists-end]
import Lean  -- necessary only for the last tactic example

section «2»
namespace Typing
axiom Ctxt : Type
axiom Term : Type
axiom Typ  : Type
axiom Typing : Ctxt → Term → Typ → Type
-- typing
notation Γ "⊢" e ":" τ => Typing Γ e τ
-- end
#check fun a b c => a ⊢ b : c
--macro Γ:term "⊢" e:term ":" τ:term : term => `(Typing $Γ $e $τ)
--
--syntax term "⊢" term ":" term : term
--macro_rules
--  | `($Γ ⊢ $e : $τ) => `(Typing $Γ $e $τ)
end Typing

-- already built into Lean
/-
-- exists
notation "∃" b "," P => Exists (fun b => P)
-- end
-/
#check ∃ x, x = x
#check ∃ (x : Nat), x = x


-- defthunk
macro "defthunk" id:ident ":=" e:term : command =>
  `(def $id := Thunk.mk (fun _ => $e))
defthunk big := mkArray 100000 true
-- end
#check big


namespace Union
abbrev Set (α : Type) := α → Prop
axiom setOf {α : Type} : (α → Prop) → Set α
axiom mem {α : Type} : α → Set α → Prop
axiom univ {α : Type} : Set α
axiom Union {α : Type} : Set (Set α) → Set α
macro:100 x:term " ∈ " s:term:99 : term => `(mem $x $s)

-- union
syntax "{" term "|" term "}" : term
macro_rules
  | `({$x ∈ $s | $p}) => `(setOf (fun $x => $x ∈ $s ∧ $p))
  | `({$b      | $p}) => `(setOf (fun $b => $p))

notation "⋃" b "," p => Union {b | p}
-- end

#check ⋃ x,              x = x
#check ⋃ (x : Set Unit), x = x
#check ⋃ x ∈ univ,       x = x


-- le
macro_rules
  | `({$x ≤ $e | $p}) => `(setOf (fun $x => $x ≤ $e ∧ $p))
-- end

#check {x ≤ 1 | x = x}
end Union


-- final example: see ./Bigop.lean
end «2»


section «3»
-- see also ./Expander.lean

-- hygiene_example
def x := 1
def e := fun (y : Nat) => x
notation "const" e => fun (x : Nat) => e
def y := const x
-- end
#check y


-- hygiene_example2
macro "m" n:ident : command => `(
  def f := 1
  macro "mm" : command => `(
    def $n := f + 1
    def f := $n + 1))
-- end
m f
mm
mm
end «3»


section «6»
-- myTac
macro "myTac" : tactic => `(intro h; exact h)
theorem triv (p : Prop) : p → p := by myTac
-- end

open Lean.Elab.Tactic
-- myTac2
def myTac2 : TacticM Unit := do
  let stx ← `(tactic| intro h; exact h)
  evalTactic stx
-- end
end «6»
\end{lstlisting}
The lack of static restrictions on the right-hand side ensures that this works just as well with custom binding notations,
even ones whose translation cannot statically be determined before substitution.
\begin{lstlisting}[linerange=union-end]
import Lean  -- necessary only for the last tactic example

section «2»
namespace Typing
axiom Ctxt : Type
axiom Term : Type
axiom Typ  : Type
axiom Typing : Ctxt → Term → Typ → Type
-- typing
notation Γ "⊢" e ":" τ => Typing Γ e τ
-- end
#check fun a b c => a ⊢ b : c
--macro Γ:term "⊢" e:term ":" τ:term : term => `(Typing $Γ $e $τ)
--
--syntax term "⊢" term ":" term : term
--macro_rules
--  | `($Γ ⊢ $e : $τ) => `(Typing $Γ $e $τ)
end Typing

-- already built into Lean
/-
-- exists
notation "∃" b "," P => Exists (fun b => P)
-- end
-/
#check ∃ x, x = x
#check ∃ (x : Nat), x = x


-- defthunk
macro "defthunk" id:ident ":=" e:term : command =>
  `(def $id := Thunk.mk (fun _ => $e))
defthunk big := mkArray 100000 true
-- end
#check big


namespace Union
abbrev Set (α : Type) := α → Prop
axiom setOf {α : Type} : (α → Prop) → Set α
axiom mem {α : Type} : α → Set α → Prop
axiom univ {α : Type} : Set α
axiom Union {α : Type} : Set (Set α) → Set α
macro:100 x:term " ∈ " s:term:99 : term => `(mem $x $s)

-- union
syntax "{" term "|" term "}" : term
macro_rules
  | `({$x ∈ $s | $p}) => `(setOf (fun $x => $x ∈ $s ∧ $p))
  | `({$b      | $p}) => `(setOf (fun $b => $p))

notation "⋃" b "," p => Union {b | p}
-- end

#check ⋃ x,              x = x
#check ⋃ (x : Set Unit), x = x
#check ⋃ x ∈ univ,       x = x


-- le
macro_rules
  | `({$x ≤ $e | $p}) => `(setOf (fun $x => $x ≤ $e ∧ $p))
-- end

#check {x ≤ 1 | x = x}
end Union


-- final example: see ./Bigop.lean
end «2»


section «3»
-- see also ./Expander.lean

-- hygiene_example
def x := 1
def e := fun (y : Nat) => x
notation "const" e => fun (x : Nat) => e
def y := const x
-- end
#check y


-- hygiene_example2
macro "m" n:ident : command => `(
  def f := 1
  macro "mm" : command => `(
    def $n := f + 1
    def f := $n + 1))
-- end
m f
mm
mm
end «3»


section «6»
-- myTac
macro "myTac" : tactic => `(intro h; exact h)
theorem triv (p : Prop) : p → p := by myTac
-- end

open Lean.Elab.Tactic
-- myTac2
def myTac2 : TacticM Unit := do
  let stx ← `(tactic| intro h; exact h)
  evalTactic stx
-- end
end «6»
\end{lstlisting}
Here we explicitly make use of the \lstinline{macro_rules} abstraction level for its convenient syntactic pattern matching syntax. \lstinline{macro_rules} are ``open'' in the sense that multiple transformers for the same \lstinline{syntax} declaration can
be defined; they are tried up to the first match, starting with the newest declaration (though this can be customized
using explicit priority annotations).
Thus the following extension will not be shadowed by the \lstinline{$b} default case above:
\begin{lstlisting}[linerange=le-end]
import Lean  -- necessary only for the last tactic example

section «2»
namespace Typing
axiom Ctxt : Type
axiom Term : Type
axiom Typ  : Type
axiom Typing : Ctxt → Term → Typ → Type
-- typing
notation Γ "⊢" e ":" τ => Typing Γ e τ
-- end
#check fun a b c => a ⊢ b : c
--macro Γ:term "⊢" e:term ":" τ:term : term => `(Typing $Γ $e $τ)
--
--syntax term "⊢" term ":" term : term
--macro_rules
--  | `($Γ ⊢ $e : $τ) => `(Typing $Γ $e $τ)
end Typing

-- already built into Lean
/-
-- exists
notation "∃" b "," P => Exists (fun b => P)
-- end
-/
#check ∃ x, x = x
#check ∃ (x : Nat), x = x


-- defthunk
macro "defthunk" id:ident ":=" e:term : command =>
  `(def $id := Thunk.mk (fun _ => $e))
defthunk big := mkArray 100000 true
-- end
#check big


namespace Union
abbrev Set (α : Type) := α → Prop
axiom setOf {α : Type} : (α → Prop) → Set α
axiom mem {α : Type} : α → Set α → Prop
axiom univ {α : Type} : Set α
axiom Union {α : Type} : Set (Set α) → Set α
macro:100 x:term " ∈ " s:term:99 : term => `(mem $x $s)

-- union
syntax "{" term "|" term "}" : term
macro_rules
  | `({$x ∈ $s | $p}) => `(setOf (fun $x => $x ∈ $s ∧ $p))
  | `({$b      | $p}) => `(setOf (fun $b => $p))

notation "⋃" b "," p => Union {b | p}
-- end

#check ⋃ x,              x = x
#check ⋃ (x : Set Unit), x = x
#check ⋃ x ∈ univ,       x = x


-- le
macro_rules
  | `({$x ≤ $e | $p}) => `(setOf (fun $x => $x ≤ $e ∧ $p))
-- end

#check {x ≤ 1 | x = x}
end Union


-- final example: see ./Bigop.lean
end «2»


section «3»
-- see also ./Expander.lean

-- hygiene_example
def x := 1
def e := fun (y : Nat) => x
notation "const" e => fun (x : Nat) => e
def y := const x
-- end
#check y


-- hygiene_example2
macro "m" n:ident : command => `(
  def f := 1
  macro "mm" : command => `(
    def $n := f + 1
    def f := $n + 1))
-- end
m f
mm
mm
end «3»


section «6»
-- myTac
macro "myTac" : tactic => `(intro h; exact h)
theorem triv (p : Prop) : p → p := by myTac
-- end

open Lean.Elab.Tactic
-- myTac2
def myTac2 : TacticM Unit := do
  let stx ← `(tactic| intro h; exact h)
  evalTactic stx
-- end
end «6»
\end{lstlisting}

As a final example, we present a partial reimplementation of the arithmetic ``bigop'' notations found\footnote{\url{https://github.com/math-comp/math-comp/blob/master/mathcomp/ssreflect/bigop.v}} in Coq's
Mathematical Components
library~\cite{mahboubi2017mathematical} such as
\begin{lstlisting}
\sum_ (i <- [0, 2, 4] | i != 2) i
\end{lstlisting}
for summing over a filtered sequence of elements. The
specific bigop notations are defined in terms of a single \verb!\big_! fold operator; however, because Coq's notation
system is unable to abstract over the indexing syntax, every specific bigop notation has to redundantly repeat every
specific index notation before delegating to \verb!\big_!. In total, the 12 index notations for \verb!\big_! are duplicated for 3 different bigops in the file.\\
\begin{minipage}{1.0\linewidth}
\begin{lstlisting}
Notation "\sum_ ( i <- r ) F"     := (\big[addn/0]_(i <- r) F).
Notation "\sum_ ( i <- r | P ) F" := (\big[addn/0]_(i <- r | P) F).
...
Notation "\prod_ ( i <- r ) F"     := (\big[muln/1]_(i <- r) F).
Notation "\prod_ ( i <- r | P ) F" := (\big[muln/1]_(i <- r | P) F).
...
\end{lstlisting}
\end{minipage}
In contrast, using our system,
we can introduce a new syntactic category for index notations, interpret it once in \verb!\big_!, and define new
bigops on top of it without any redundancy.
\begin{lstlisting}[linerange=declare_syntax_cat-end,belowskip=0pt]
namespace Bigop

section Preliminaries

def Seq (α : Type) := List α

def BigBody (β α) :=  α × (β → β → β) × Bool × β

def applyBig (body : BigBody β α) (x : β) : β :=
   let (_, op, b, v) := body
   if b then op v x else x

def reducebig (idx : β) (r : Seq α) (body : α → BigBody β α) : β :=
   r.foldr (applyBig ∘ body) idx

def bigop := @reducebig

def iota : Nat → Nat → List Nat
   | m, 0   => []
   | m, n+1 => m :: iota (m+1) n

def index_iota (m n : Nat) := iota m (n - m)

class Enumerable (α : Type) where
   elems : List α

instance : Enumerable Bool where
   elems := [false, true]

instance [Enumerable α] [Enumerable β]: Enumerable (α × β) where
  elems := Enumerable.elems.bind fun a => Enumerable.elems.map fun b => (a, b)

def finElemsAux (n : Nat) : (i : Nat) → i < n → List (Fin n)
   | 0,   h => [⟨0, h⟩]
   | i+1, h => ⟨i+1, h⟩ :: finElemsAux n i (Nat.ltOfSuccLt h)

def finElems : (n : Nat) → List (Fin n)
   | 0     => []
   | (n+1) => finElemsAux (n+1) n (Nat.ltSuccSelf n)

instance : Enumerable (Fin n) where
  elems := (finElems n).reverse

instance : OfNat (Fin (Nat.succ n)) m where
  ofNat := Fin.ofNat m

end Preliminaries

-- Declare a new syntax category for "indexing" big operators
-- declare_syntax_cat
declare_syntax_cat index
syntax ident "<-" term : index
syntax ident "<-" term "|" term : index
-- end
syntax term:51 "≤" ident "<" term : index
syntax term:51 "≤" ident "<" term "|" term : index
-- Primitive notation for big operators
syntax "\\big_" "[" term "," term "]" "(" index ")" term : term

-- We define how to expand `\big` with the different kinds of index
macro_rules
  | `(\big_ [$op, $idx] ($i:ident <- $r) $F) => `(bigop $idx $r (fun $i => ($i, $op, true, $F)))
  | `(\big_ [$op, $idx] ($i:ident <- $r | $p) $F) => `(bigop $idx $r (fun $i => ($i, $op, $p, $F)))
  | `(\big_ [$op, $idx] ($lower:term ≤ $i:ident < $upper) $F) => `(bigop $idx (index_iota $lower $upper) (fun $i => ($i, $op, true, $F)))
  | `(\big_ [$op, $idx] ($lower:term ≤ $i:ident < $upper | $p) $F) => `(bigop $idx (index_iota $lower $upper) (fun $i => ($i, $op, $p, $F)))

-- Sum
macro "Σ" "(" idx:index ")" F:term : term =>
  `(\big_ [Add.add, 0] ($idx) $F)
-- end

-- We can already use `Σ` with the different kinds of index.
#check Σ (i <- [0, 2, 4] | i != 2) i
#check Σ (10 ≤ i < 20 | i != 5) i+1
#check Σ (10 ≤ i < 20) i+1

-- Define `Π`
-- Prod
macro "Π" "(" idx:index ")" F:term : term =>
  `(\big_ [Mul.mul, 1] ($idx) $F)
-- end

-- The examples above now also work for `Π`
#check Π (i <- [0, 2, 4] | i != 2) i
#check Π (10 ≤ i < 20 | i != 5) i+1
#check Π (10 ≤ i < 20) i+1

-- We can extend our grammar for the syntax category `index`.
syntax ident "|" term : index
syntax ident ":" term : index
syntax ident ":" term "|" term : index
-- Add new rules
macro_rules
  | `(\big_ [$op, $idx] ($i:ident : $type) $F) => `(bigop $idx (Enumerable.elems (α := $type)) (fun $i => ($i, $op, true, $F)))
  | `(\big_ [$op, $idx] ($i:ident : $type | $p) $F) => `(bigop $idx (Enumerable.elems (α := $type)) (fun $i => ($i, $op, $p, $F)))
  | `(\big_ [$op, $idx] ($i:ident | $p) $F) => `(bigop $idx Enumerable.elems (fun $i => ($i, $op, $p, $F)))

-- The new syntax is immediately available for all big operators that we have defined
def myPred (x : Fin 10) : Bool := true
#check Σ (i : Fin 10) i+1
#check Σ (i : Fin 10 | i != 2) i+1
#check Σ (i | myPred i) i+i
#check Π (i : Fin 10) i+1
#check Π (i : Fin 10 | i != 2) i+1
#check Π (i | myPred i) i+i

-- We can easily create alternative syntax for any big operator.
macro "Σ" idx:index "," F:term : term =>
  `(Σ ($idx) $F)

#check Σ 10 ≤ i < 20, i+1

-- Finally, we create a command for automating the generation of big operators.
syntax "def_bigop" str term:max term:max : command
-- Antiquotations can be nested as in `$$F`, which expands to `$F`, i.e. the value of
-- `F` is inserted only in the second expansion, the expansion of the new macro `$head`.
macro_rules
  | `(def_bigop $head $op $unit) =>
    `(macro $head:strLit "(" idx:index ")" F:term : term => `(\big_ [$op, $unit] ($$idx) $$F))

def_bigop "SUM" Nat.add 0
#check SUM (i <- [0, 1, 2]) i+1

end Bigop
\end{lstlisting}
\begin{lstlisting}[aboveskip=0pt,belowskip=0pt,showlines=true]
...

\end{lstlisting}
\begin{lstlisting}[linerange=Sum-end,aboveskip=0pt]
namespace Bigop

section Preliminaries

def Seq (α : Type) := List α

def BigBody (β α) :=  α × (β → β → β) × Bool × β

def applyBig (body : BigBody β α) (x : β) : β :=
   let (_, op, b, v) := body
   if b then op v x else x

def reducebig (idx : β) (r : Seq α) (body : α → BigBody β α) : β :=
   r.foldr (applyBig ∘ body) idx

def bigop := @reducebig

def iota : Nat → Nat → List Nat
   | m, 0   => []
   | m, n+1 => m :: iota (m+1) n

def index_iota (m n : Nat) := iota m (n - m)

class Enumerable (α : Type) where
   elems : List α

instance : Enumerable Bool where
   elems := [false, true]

instance [Enumerable α] [Enumerable β]: Enumerable (α × β) where
  elems := Enumerable.elems.bind fun a => Enumerable.elems.map fun b => (a, b)

def finElemsAux (n : Nat) : (i : Nat) → i < n → List (Fin n)
   | 0,   h => [⟨0, h⟩]
   | i+1, h => ⟨i+1, h⟩ :: finElemsAux n i (Nat.ltOfSuccLt h)

def finElems : (n : Nat) → List (Fin n)
   | 0     => []
   | (n+1) => finElemsAux (n+1) n (Nat.ltSuccSelf n)

instance : Enumerable (Fin n) where
  elems := (finElems n).reverse

instance : OfNat (Fin (Nat.succ n)) m where
  ofNat := Fin.ofNat m

end Preliminaries

-- Declare a new syntax category for "indexing" big operators
-- declare_syntax_cat
declare_syntax_cat index
syntax ident "<-" term : index
syntax ident "<-" term "|" term : index
-- end
syntax term:51 "≤" ident "<" term : index
syntax term:51 "≤" ident "<" term "|" term : index
-- Primitive notation for big operators
syntax "\\big_" "[" term "," term "]" "(" index ")" term : term

-- We define how to expand `\big` with the different kinds of index
macro_rules
  | `(\big_ [$op, $idx] ($i:ident <- $r) $F) => `(bigop $idx $r (fun $i => ($i, $op, true, $F)))
  | `(\big_ [$op, $idx] ($i:ident <- $r | $p) $F) => `(bigop $idx $r (fun $i => ($i, $op, $p, $F)))
  | `(\big_ [$op, $idx] ($lower:term ≤ $i:ident < $upper) $F) => `(bigop $idx (index_iota $lower $upper) (fun $i => ($i, $op, true, $F)))
  | `(\big_ [$op, $idx] ($lower:term ≤ $i:ident < $upper | $p) $F) => `(bigop $idx (index_iota $lower $upper) (fun $i => ($i, $op, $p, $F)))

-- Sum
macro "Σ" "(" idx:index ")" F:term : term =>
  `(\big_ [Add.add, 0] ($idx) $F)
-- end

-- We can already use `Σ` with the different kinds of index.
#check Σ (i <- [0, 2, 4] | i != 2) i
#check Σ (10 ≤ i < 20 | i != 5) i+1
#check Σ (10 ≤ i < 20) i+1

-- Define `Π`
-- Prod
macro "Π" "(" idx:index ")" F:term : term =>
  `(\big_ [Mul.mul, 1] ($idx) $F)
-- end

-- The examples above now also work for `Π`
#check Π (i <- [0, 2, 4] | i != 2) i
#check Π (10 ≤ i < 20 | i != 5) i+1
#check Π (10 ≤ i < 20) i+1

-- We can extend our grammar for the syntax category `index`.
syntax ident "|" term : index
syntax ident ":" term : index
syntax ident ":" term "|" term : index
-- Add new rules
macro_rules
  | `(\big_ [$op, $idx] ($i:ident : $type) $F) => `(bigop $idx (Enumerable.elems (α := $type)) (fun $i => ($i, $op, true, $F)))
  | `(\big_ [$op, $idx] ($i:ident : $type | $p) $F) => `(bigop $idx (Enumerable.elems (α := $type)) (fun $i => ($i, $op, $p, $F)))
  | `(\big_ [$op, $idx] ($i:ident | $p) $F) => `(bigop $idx Enumerable.elems (fun $i => ($i, $op, $p, $F)))

-- The new syntax is immediately available for all big operators that we have defined
def myPred (x : Fin 10) : Bool := true
#check Σ (i : Fin 10) i+1
#check Σ (i : Fin 10 | i != 2) i+1
#check Σ (i | myPred i) i+i
#check Π (i : Fin 10) i+1
#check Π (i : Fin 10 | i != 2) i+1
#check Π (i | myPred i) i+i

-- We can easily create alternative syntax for any big operator.
macro "Σ" idx:index "," F:term : term =>
  `(Σ ($idx) $F)

#check Σ 10 ≤ i < 20, i+1

-- Finally, we create a command for automating the generation of big operators.
syntax "def_bigop" str term:max term:max : command
-- Antiquotations can be nested as in `$$F`, which expands to `$F`, i.e. the value of
-- `F` is inserted only in the second expansion, the expansion of the new macro `$head`.
macro_rules
  | `(def_bigop $head $op $unit) =>
    `(macro $head:strLit "(" idx:index ")" F:term : term => `(\big_ [$op, $unit] ($$idx) $$F))

def_bigop "SUM" Nat.add 0
#check SUM (i <- [0, 1, 2]) i+1

end Bigop
\end{lstlisting}
\begin{lstlisting}[linerange=Prod-end,aboveskip=0pt]
namespace Bigop

section Preliminaries

def Seq (α : Type) := List α

def BigBody (β α) :=  α × (β → β → β) × Bool × β

def applyBig (body : BigBody β α) (x : β) : β :=
   let (_, op, b, v) := body
   if b then op v x else x

def reducebig (idx : β) (r : Seq α) (body : α → BigBody β α) : β :=
   r.foldr (applyBig ∘ body) idx

def bigop := @reducebig

def iota : Nat → Nat → List Nat
   | m, 0   => []
   | m, n+1 => m :: iota (m+1) n

def index_iota (m n : Nat) := iota m (n - m)

class Enumerable (α : Type) where
   elems : List α

instance : Enumerable Bool where
   elems := [false, true]

instance [Enumerable α] [Enumerable β]: Enumerable (α × β) where
  elems := Enumerable.elems.bind fun a => Enumerable.elems.map fun b => (a, b)

def finElemsAux (n : Nat) : (i : Nat) → i < n → List (Fin n)
   | 0,   h => [⟨0, h⟩]
   | i+1, h => ⟨i+1, h⟩ :: finElemsAux n i (Nat.ltOfSuccLt h)

def finElems : (n : Nat) → List (Fin n)
   | 0     => []
   | (n+1) => finElemsAux (n+1) n (Nat.ltSuccSelf n)

instance : Enumerable (Fin n) where
  elems := (finElems n).reverse

instance : OfNat (Fin (Nat.succ n)) m where
  ofNat := Fin.ofNat m

end Preliminaries

-- Declare a new syntax category for "indexing" big operators
-- declare_syntax_cat
declare_syntax_cat index
syntax ident "<-" term : index
syntax ident "<-" term "|" term : index
-- end
syntax term:51 "≤" ident "<" term : index
syntax term:51 "≤" ident "<" term "|" term : index
-- Primitive notation for big operators
syntax "\\big_" "[" term "," term "]" "(" index ")" term : term

-- We define how to expand `\big` with the different kinds of index
macro_rules
  | `(\big_ [$op, $idx] ($i:ident <- $r) $F) => `(bigop $idx $r (fun $i => ($i, $op, true, $F)))
  | `(\big_ [$op, $idx] ($i:ident <- $r | $p) $F) => `(bigop $idx $r (fun $i => ($i, $op, $p, $F)))
  | `(\big_ [$op, $idx] ($lower:term ≤ $i:ident < $upper) $F) => `(bigop $idx (index_iota $lower $upper) (fun $i => ($i, $op, true, $F)))
  | `(\big_ [$op, $idx] ($lower:term ≤ $i:ident < $upper | $p) $F) => `(bigop $idx (index_iota $lower $upper) (fun $i => ($i, $op, $p, $F)))

-- Sum
macro "Σ" "(" idx:index ")" F:term : term =>
  `(\big_ [Add.add, 0] ($idx) $F)
-- end

-- We can already use `Σ` with the different kinds of index.
#check Σ (i <- [0, 2, 4] | i != 2) i
#check Σ (10 ≤ i < 20 | i != 5) i+1
#check Σ (10 ≤ i < 20) i+1

-- Define `Π`
-- Prod
macro "Π" "(" idx:index ")" F:term : term =>
  `(\big_ [Mul.mul, 1] ($idx) $F)
-- end

-- The examples above now also work for `Π`
#check Π (i <- [0, 2, 4] | i != 2) i
#check Π (10 ≤ i < 20 | i != 5) i+1
#check Π (10 ≤ i < 20) i+1

-- We can extend our grammar for the syntax category `index`.
syntax ident "|" term : index
syntax ident ":" term : index
syntax ident ":" term "|" term : index
-- Add new rules
macro_rules
  | `(\big_ [$op, $idx] ($i:ident : $type) $F) => `(bigop $idx (Enumerable.elems (α := $type)) (fun $i => ($i, $op, true, $F)))
  | `(\big_ [$op, $idx] ($i:ident : $type | $p) $F) => `(bigop $idx (Enumerable.elems (α := $type)) (fun $i => ($i, $op, $p, $F)))
  | `(\big_ [$op, $idx] ($i:ident | $p) $F) => `(bigop $idx Enumerable.elems (fun $i => ($i, $op, $p, $F)))

-- The new syntax is immediately available for all big operators that we have defined
def myPred (x : Fin 10) : Bool := true
#check Σ (i : Fin 10) i+1
#check Σ (i : Fin 10 | i != 2) i+1
#check Σ (i | myPred i) i+i
#check Π (i : Fin 10) i+1
#check Π (i : Fin 10 | i != 2) i+1
#check Π (i | myPred i) i+i

-- We can easily create alternative syntax for any big operator.
macro "Σ" idx:index "," F:term : term =>
  `(Σ ($idx) $F)

#check Σ 10 ≤ i < 20, i+1

-- Finally, we create a command for automating the generation of big operators.
syntax "def_bigop" str term:max term:max : command
-- Antiquotations can be nested as in `$$F`, which expands to `$F`, i.e. the value of
-- `F` is inserted only in the second expansion, the expansion of the new macro `$head`.
macro_rules
  | `(def_bigop $head $op $unit) =>
    `(macro $head:strLit "(" idx:index ")" F:term : term => `(\big_ [$op, $unit] ($$idx) $$F))

def_bigop "SUM" Nat.add 0
#check SUM (i <- [0, 1, 2]) i+1

end Bigop
\end{lstlisting}
The full example is included in the supplement.
We reiterate that this is not merely a showcase of a parsing extension, but that abstracting over binding syntax in this manner is fundamentally incompatible with any static approach of ensuring hygiene.
The dynamic nature of Scheme-like macros allows us to always apply such factorings without being burdened by static restrictions while still preserving hygiene.

\section{Hygiene Algorithm}
\label{sec:hygge}

In this section, we will give a mostly self-contained description of our algorithm for automatic hygiene applied to a
simple recursive macro expander; we postpone comparisons to existing hygiene algorithms to \cref{sec:related}.

Hygiene issues occur when transformations such as macro expansions lead to an unexpected capture (rebinding) of
identifiers. For example, we would expect the notation
\begin{lstlisting}
notation "const" e => fun x => e
\end{lstlisting}
to always produce a constant function regardless of the specific \lstinline{e}.
We would not expect the term \lstinline{const x} to be the identity function because intuitively there is no
\verb!x! in scope at the argument position of \lstinline{const}; that the implementation of the macro
makes use of the name internally should be of no concern to the macro user.

Thus hygiene issues can also be described as a \emph{confusion of scopes} when syntax parts are removed from their
original context and inserted into new contexts, which makes name resolution strictly after macro expansion (such as in a compiler preceded
by a preprocessor) futile. Instead we need to track scopes \emph{as metadata} before and during macro expansion so as
not to lose information about the original context of identifiers. Specifically,
\begin{enumerate}
\item when an identifier captured in a syntax quotation matches one or more\footnote{Lean allows overloaded
        top-level bindings whereas local bindings are shadowing.}
        top-level symbols, the identifier is annotated with a list of these symbols as
  \emph{top-level} scopes to
  preserve its \emph{extra-macro} context (which, because of the lack of local macros, can only contain top-level
  bindings), and
\item when a macro is expanded, all identifiers freshly introduced by the expansion are annotated with a new \emph{macro} scope
  to preserve the \emph{intra-macro} context.
  In particular, different expansions of the same macro introduce different annotations.
  Macro scopes are appended to a list, i.e.\ ordered by expansion time.
  This full ``history of expansions'' is necessary to treat macro-producing macros correctly, as we shall see in \cref{sec:hygge:ex}.
\end{enumerate}
Thus, the expansion of the above term \lstinline{const x} should be (an equivalent of)
\lstinline{fun x.1 => x}
where \verb!1! is a fresh macro scope appended to the macro-introduced \verb!x!, preventing it from capturing the
\verb!x! from the original input. In general, we will present hygienic identifiers in the following as \texttt{n.msc$_1$.msc$_2$.$\ldots$.msc$_n$\{tsc$_1$,$\ldots$,tsc$_n$\}}
where \verb!n! is the original name, \verb!msc! are macro scopes, and \verb!tsc! top-level scopes, eliding the braces if
there are no top-level scopes as in the example above. We use the dot
notation to suggest both the ordered nature of macro scopes and their eventual implementation in \cref{sec:impl}.
We will now describe how to implement these operations in a standard macro expander.


\subsection{Expansion Algorithm}
\label{sec:algo}

A Scheme-style macro expander takes a syntax tree as input and produces a fully expanded tree, that is, where all macro uses have been reduced to \emph{core forms} that cannot be described as macros and are instead handled by the later stages, such as an elaborator.
The expander should furthermore rename bindings where necessary to avoid hygiene issues such that later stages do not have to know anything about the implementation of hygiene, or indeed that it was applied at all.

Given a \emph{global context} (a set of symbols), our expansion algorithm does so by a conventional top-down expansion, keeping
track of an initially-empty \emph{local context} (another set of symbols). 
When a binding core form is encountered, the local context is extended with the bound symbol(s); existing top-level
scopes on bindings are discarded at this step since they are only meaningful on references.
Thus we will formally define a symbol as an identifier together with a list of macro scopes, such as \verb!x.1! above.
As we shall see in \cref{sec:impl}, this definition of symbol is covered by the existing one in Lean, so later stages indeed do not have to concern themselves with it.

When a reference (another core form), i.e.\ an identifier not
in binding position, is encountered, it is resolved according to the following rules:
\begin{enumerate}
\item If the local context has an entry for the same symbol, the reference binds to
  the corresponding local binding; any top-level scopes are again discarded.
\item Otherwise, if the identifier is annotated with one or more top-level scopes or matches one or more symbols in the
  global context, it binds to all of these (to be disambiguated by the elaborator).
%
\item Otherwise, the identifier is unbound and an error is generated.
\end{enumerate}

In the common incremental compilation mode of ITPs, every command is fully processed before subsequent commands.
Thus, an expander for such a system will never extend the global context by itself, but pass the fully expanded command
to the next compilation stage before being called again with the next command's unexpanded syntax tree and a possibly
extended global context.

Notably, our expander does not introduce macro scopes by itself, either, much in contrast to other expansion algorithms. We
instead delegate this task to the macro's implementation, though in a completely transparent way for all pattern-based and for
many procedural macros. We claim that a macro should in fact be interpreted as an \emph{effectful} computation since two
expansions of the same identifier-introducing macro should not return the same syntax tree to avoid unhygienic
interactions between them. Thus,
as a \emph{side effect}, it should apply a fresh macro scope to each newly introduced identifier. In particular, a syntax quotation
should not merely be seen as a datum, but as an effectful value that obtains and applies this fresh scope to
all the identifiers captured by it to immediately ensure hygiene for pattern-based macros. Procedural macros producing
identifiers not originating from syntax quotations might need to obtain and make use of the fresh macro scope
explicitly. Note that forgoing to do so is not sufficient to reliably implement \emph{anaphoric} or other hygiene-bending macros
  that make an identifier (conventionally \lstinline{it} in Lisp languages) available in the scope of the macro caller, as discussed in \citet{barzilay2011keeping}.
  Instead, we believe that the correct translation of anaphoric macros to Lean is to change such identifiers to \emph{keywords} that
  do not participate in hygiene at all, analogous to the \emph{syntax parameters} of \citet{barzilay2011keeping}.
  An example for this is the \lstinline{this} keyword introduced by tactics such as \lstinline{have}
  that can be used to refer to the just-proved fact.\footnote{\url{https://github.com/leanprover/lean4/blob/6d0c91c/src/Init/Notation.lean\#L204-L207}}

We give a specific monad-based~\cite{moggi1991notions}
implementation of effectful syntax quotations as a regular macro in \cref{sec:impl}.
For the remainder of this section, we will simply assume that macros have been implemented in this way and observe their interaction with the expansion algorithm described above.

\subsection{Examples}
\label{sec:hygge:ex}

Given the following input,
\begin{lstlisting}
def x := 1
def e := fun y => x
notation "const" e => fun x => e
def y := const x
\end{lstlisting}
a Lean-like system using the presented expansion algorithm should incrementally parse, expand, and elaborate each declaration before advancing to the next one.
For a first, trivial example, let us
focus on the expansion of the second line. At this point, the global context contains the symbol \verb!x! (plus any
default imports that we will ignore here). Descending into the right-hand side of the definition, the expander first adds \verb!y! to the
local context. The reference \verb!x! does not match any local definitions, so it binds to the matching top-level
definition.

In the next line, the built-in \lstinline{notation} macro expands to the definitions
\begin{lstlisting}
syntax "const" term : term
macro_rules
  | `(const $e) => `(fun x => $e)
\end{lstlisting}
When a top-level macro application unfolds to multiple declarations, we expand and elaborate these
incrementally as well to ensure that declarations are in the global context of subsequent declarations from the same expansion. When
recursively expanding the \lstinline{macro_rules} declaration (we will assume for this example that \lstinline{macro_rules} itself is
a core form) in the global context \verb!{x, e}!, we first visit the syntax quotation on the left-hand side. The
identifier \verb!e! inside of it is in an antiquotation and thus not captured by the quotation. It is in binding
position for the right-hand side, so we add \verb!e! to the local context.
Visiting the right-hand side, we find the quotation-captured identifier \verb!x! and annotate it with the matching
top-level definition of the same name; we do not yet know that it is in a binding position. When visiting the reference
\verb!e!, we see that it matches a local binding and do not add top-level scopes.
\begin{lstlisting}
macro_rules
  | `(const $e) => `(fun x{x} => $e)
\end{lstlisting}
Visiting the last line
\begin{lstlisting}
def y := const x
\end{lstlisting}
with the global context \verb!{x, e}!, we descend into the right-hand side. We
expand the \lstinline{const} macro given a fresh macro scope \verb!1!, which is applied to any captured identifiers.
\begin{lstlisting}
def y := fun x.1{x} => x
\end{lstlisting}
We add the symbol \verb!x.1! (discarding the top-level scope \verb!x!) to the
local context and finally visit the reference \verb!x!. The reference does not match the local binding \verb!x.1! but
does match the top-level binding \verb!x!, so it binds to the latter.
\begin{lstlisting}
def y := fun x.1 => x
\end{lstlisting}

Now let us briefly look at a more complex macro-macro example demonstrating use of the macro scopes stack:
\begin{lstlisting}[linerange=hygiene_example2-end]
import Lean  -- necessary only for the last tactic example

section «2»
namespace Typing
axiom Ctxt : Type
axiom Term : Type
axiom Typ  : Type
axiom Typing : Ctxt → Term → Typ → Type
-- typing
notation Γ "⊢" e ":" τ => Typing Γ e τ
-- end
#check fun a b c => a ⊢ b : c
--macro Γ:term "⊢" e:term ":" τ:term : term => `(Typing $Γ $e $τ)
--
--syntax term "⊢" term ":" term : term
--macro_rules
--  | `($Γ ⊢ $e : $τ) => `(Typing $Γ $e $τ)
end Typing

-- already built into Lean
/-
-- exists
notation "∃" b "," P => Exists (fun b => P)
-- end
-/
#check ∃ x, x = x
#check ∃ (x : Nat), x = x


-- defthunk
macro "defthunk" id:ident ":=" e:term : command =>
  `(def $id := Thunk.mk (fun _ => $e))
defthunk big := mkArray 100000 true
-- end
#check big


namespace Union
abbrev Set (α : Type) := α → Prop
axiom setOf {α : Type} : (α → Prop) → Set α
axiom mem {α : Type} : α → Set α → Prop
axiom univ {α : Type} : Set α
axiom Union {α : Type} : Set (Set α) → Set α
macro:100 x:term " ∈ " s:term:99 : term => `(mem $x $s)

-- union
syntax "{" term "|" term "}" : term
macro_rules
  | `({$x ∈ $s | $p}) => `(setOf (fun $x => $x ∈ $s ∧ $p))
  | `({$b      | $p}) => `(setOf (fun $b => $p))

notation "⋃" b "," p => Union {b | p}
-- end

#check ⋃ x,              x = x
#check ⋃ (x : Set Unit), x = x
#check ⋃ x ∈ univ,       x = x


-- le
macro_rules
  | `({$x ≤ $e | $p}) => `(setOf (fun $x => $x ≤ $e ∧ $p))
-- end

#check {x ≤ 1 | x = x}
end Union


-- final example: see ./Bigop.lean
end «2»


section «3»
-- see also ./Expander.lean

-- hygiene_example
def x := 1
def e := fun (y : Nat) => x
notation "const" e => fun (x : Nat) => e
def y := const x
-- end
#check y


-- hygiene_example2
macro "m" n:ident : command => `(
  def f := 1
  macro "mm" : command => `(
    def $n := f + 1
    def f := $n + 1))
-- end
m f
mm
mm
end «3»


section «6»
-- myTac
macro "myTac" : tactic => `(intro h; exact h)
theorem triv (p : Prop) : p → p := by myTac
-- end

open Lean.Elab.Tactic
-- myTac2
def myTac2 : TacticM Unit := do
  let stx ← `(tactic| intro h; exact h)
  evalTactic stx
-- end
end «6»
\end{lstlisting}
If we use this macro as in \lstinline{m f}, we apply a fresh macro scope \verb!1! to all captured identifiers, then incrementally process
the two new declarations.
\begin{lstlisting}
def f.1 := 1
macro "mm" : command => `(
  def f := f.1{f.1} + 1
  def f.1{f.1} := f + 1)
\end{lstlisting}
If we use the new macro \lstinline{mm}, we apply one more macro scope \verb!2!.
\begin{lstlisting}
def f.2 := f.1.2{f.1} + 1
def f.1.2{f.1} := f.2 + 1
\end{lstlisting}
When processing these new definitions, we see that the scopes ensure the expected name resolution.
\begin{lstlisting}
def f.1 := 1
...
def f.2 := f.1 + 1
def f.1.2 := f.2 + 1
\end{lstlisting}
In particular, we now
have global declarations \lstinline{f.1}, \lstinline{f.2}, and \lstinline{f.1.2} that show that storing only a single
macro scope would have led to a collision.


%
%

\section{Implementation}
\label{sec:impl}

Syntax objects in Lean 4 are represented as an inductive type of \emph{nodes} (or nonterminals), \emph{atoms} (or
terminals), and, as a special case of terminals, \emph{identifiers}.
\begin{lstlisting}
inductive Syntax where
  | node  (kind : Name) (args : Array Syntax)
  | atom  (info : SourceInfo) (val : String)
  | ident (info : SourceInfo) (rawVal : String) (val : Name) (preresolved : List (Nat × List String))
  | missing
\end{lstlisting}
An additional constructor represents \emph{missing} parts from syntax error recovery.
Atoms and identifiers are annotated with source location metadata unless generated by a macro.
Identifiers carry macro scopes inline in their \verb!Name! while top-level scopes are held in a
separate list. The additional \verb!Nat! is an implementation detail of Lean's hierarchical name resolution.

The type \verb!Name! of hierarchical names precedes the implementation of the macro system and is used throughout Lean's
implementation for referring to (namespaced) symbols.
\begin{lstlisting}
inductive Name where
  | anonymous
  | str (base : Name) (s : String)
  | num (base : Name) (n : Nat)
\end{lstlisting}
The syntax \lstinline{`a.b} is a literal of type \lstinline{Name} for use in meta-programs. The numeric part of
\lstinline{Name} is not accessible from the surface syntax and reserved for internal names; similar designs are found in
other ITPs. By reusing \lstinline{Name} for storing
macro scopes, but not top-level scopes, we ensure that the new definition of \emph{symbol} from \cref{sec:algo}
coincides with the existing Lean type and no changes to the implementation of the local or global context are necessary
for adopting the macro system.

\begin{figure}[tb]
\begin{lstlisting}[linerange=expand-end,numbers=left,firstnumber=0,belowskip=0pt]
import Lean

namespace Lean
namespace Expander

open Lean.Syntax

-- Result of name resolution. As in the paper, we will ignore the second component here.
abbrev NameRes := Name × List String
-- We model the global context more precisely as a mapping from symbols to qualified symbols,
-- e.g. (`a ↦ [`ns1.a, `ns2.a])
abbrev GlobalContext := Name → List NameRes

-- the simplified transformer monad
structure TransformerContext where
  gctx : GlobalContext
  currMacroScope : MacroScope

abbrev TransformerM := ReaderT TransformerContext Id
abbrev Transformer := Syntax → TransformerM Syntax

-- support syntax quotations in transformers
instance : MonadQuotation TransformerM where
  getCurrMacroScope   := do let ctx ← read; pure ctx.currMacroScope
  -- dummy impls, unused
  withFreshMacroScope := fun x => x
  getRef              := pure Syntax.missing
  withRef             := fun _ x => x
  -- The actual implementation also adds the current module name to macro scopes for global uniqueness,
  -- which we can ignore in this single-file example.
  getMainModule       := pure `Expander

-- the expander extension of the transformer monad
structure ExpanderContext extends TransformerContext where
  lctx : NameSet
  macros : Name → Option Transformer

abbrev ExpanderM := ReaderT ExpanderContext <| StateT MacroScope <| ExceptT String <| Id

instance MonadQuotation : MonadQuotation ExpanderM where
  getCurrMacroScope   := do let ctx ← read; pure ctx.currMacroScope
  withFreshMacroScope := fun x => do
    let fresh ← modifyGet (fun n => (n, n + 1))
    withReader (fun ctx => { ctx with currMacroScope := fresh }) x
  getMainModule       := pure `Expander
  -- dummy impls, unused
  getRef              := pure Syntax.missing
  withRef             := fun _ x => x

-- implicitly coerce transformer monad into expander monad
instance : Coe (TransformerM α) (ExpanderM α) where
  coe t := fun ctx => t ctx.toTransformerContext

-- simplified: ignore the module name parameter
def addMacroScope (n : Name) (scp : MacroScope) : Name :=
  Lean.addMacroScope `Expander n scp

def getGlobalContext : TransformerM GlobalContext := do
  return (← read).gctx

def getLocalContext : ExpanderM NameSet := do
  return (← read).lctx

def resolve (gctx : GlobalContext) (n : Name) : List NameRes :=
  gctx n

-- slightly more meaningful name
def getIdentVal : Syntax → Name := Syntax.getId

def getPreresolved : Syntax → List NameRes
  | Syntax.ident (preresolved := preresolved) .. => preresolved
  | _                                            => []

def mkOverloadedIds (cs : List NameRes) : Syntax :=
  Syntax.node choiceKind (cs.toArray.map (mkIdent ∘ Prod.fst))

def withLocal (l : Name) : ExpanderM Syntax → ExpanderM Syntax :=
  withReader (fun ctx => { ctx with lctx := ctx.lctx.insert l })

def getTransformerFor (k : SyntaxNodeKind) : ExpanderM (Syntax → ExpanderM Syntax) := do
  match (← read).macros k with
  | some t => return fun stx => t stx
  | none   => throw ("unknown macro " ++ toString k)

-- slightly simplified from the actual implementation
partial def getAntiquotationIds (stx : Syntax) : ExpanderM (Array Syntax) := do
  let mut ids := #[]
  for stx in stx.topDown do
    if (isAntiquot stx || isTokenAntiquot stx) && !isEscapedAntiquot stx then
      let anti := getAntiquotTerm stx
      if anti.isIdent then ids := ids.push anti
      else throw "complex antiquotation not allowed here"
  return ids

-- Get all pattern vars (as `Syntax.ident`s) in `stx`
partial def getPatternVars (stx : Syntax) : ExpanderM (Array Syntax) :=
  if stx.isQuot then
    getAntiquotationIds stx
  else match stx with
    | `(_)            => #[]
    | `($id:ident)    => #[id]
    | `($id:ident@$e) => do (← getPatternVars e).push id
    | _               => throw "unsupported pattern in syntax match"

-- expand
partial def expand : Syntax → ExpanderM Syntax
  | `($id:ident) => do
    let val : Name := getIdentVal id
    let gctx ← getGlobalContext
    let lctx ← getLocalContext
    if lctx.contains val then
      pure (mkIdent val)
    else match resolve gctx val ++ getPreresolved id with
      | []        => throw ("unknown identifier " ++ toString val)
      | [(id, _)] => pure (mkIdent id)
      | ids       => pure (mkOverloadedIds ids)
  | `(fun ($id : $ty) => $e) => do
    let val := getIdentVal id
    let ty ← expand ty
    let e ← withLocal val (expand e)
    `(fun ($(mkIdent val) : $ty) => $e)
-- end
  -- more core forms
  | `(fun $id:ident => $e) => do
    let e ← withLocal (getIdentVal id) (expand e)
    `(fun $id:ident => $e)
  | `($num:numLit) => `($num:numLit)
  | `($str:strLit) => `($str:strLit)
  | `($n:quotedName) => `($n:quotedName)
  | `($fn $args*) => do
    let fn ← expand fn
    let args ← args.mapM expand
    `($fn $args*)
  | `(def $id := $e) => do
    let e ← expand e
    `(def $id := $e)
  -- syntax: keep as-is
  | `(syntax $[(name := $n)]? $[(priority := $prio)]? $[$args:stx]* : $kind) => `(syntax $[(name := $n)]? $[(priority := $prio)]? $[$args:stx]* : $kind)
  -- macro_rules: expand rhs (but not lhs) to exercise syntax quotation macro
  | `(macro_rules | $lhs => $rhs) => do
    let vars ← getPatternVars lhs
    let rhs ← vars.foldr (fun var ex => withLocal var.getId ex) (expand rhs)
    `(macro_rules | $lhs => $rhs)
  -- we will ignore double-backtick quotations generated by `notation` for this example
  | `(``($e)) => `(`($e))
  | stx => do
    -- expansion consists of multiple commands => yield and get called back per command
    if stx.isOfKind nullKind then pure stx else do
    let t ← getTransformerFor stx.getKind
    let stx ← withFreshMacroScope (t stx)
    expand stx

open Lean.Elab.Term.Quotation
-- quoteSyntax
partial def quoteSyntax : Syntax → TransformerM Syntax
  | Syntax.ident info rawVal val preresolved => do
    let gctx ← getGlobalContext
    let preresolved := resolve gctx val ++ preresolved
    `(Syntax.ident SourceInfo.none $(quote rawVal)
        (addMacroScope $(quote val) msc) $(quote preresolved))
  | stx@(Syntax.node k args) =>
    if isAntiquot stx then pure (getAntiquotTerm stx)
    else do
      let args ← args.mapM quoteSyntax
      `(Syntax.node $(quote k) $(quote args))
  | Syntax.atom info val => `(Syntax.atom SourceInfo.none $(quote val))
  | Syntax.missing => pure Syntax.missing

def expandStxQuot (stx : Syntax) : TransformerM Syntax := do
  let stx ← quoteSyntax (stx.getArg 1)
  `(do msc ← getCurrMacroScope; pure $stx)
-- end

-- two more, simple macros
def expandDo : Transformer
  | `(do $id:ident ← $val:term; $body:term) => `(Bind.bind $val (fun $id:ident => $body))
  | _                                  => pure Syntax.missing

def expandParen : Transformer
  | `(($e)) => pure e
  | _       => pure Syntax.missing

-- custom Syntax pretty printer for our core forms that uses the paper's notation for hygienic identifiers
def ppIdent (n : Name) : Format :=
  let v := extractMacroScopes n
  fmt <| v.scopes.foldl Name.mkNum v.name

-- flip to make output more readable
def hideMacroRulesRhs := false

open Std.Format
partial def pp : Syntax → Format
  | `($id:ident) => match getPreresolved id with
    | [] => ppIdent id.getId
    | ps => ppIdent id.getId ++ bracket "{" (joinSep (ps.map (fmt ∘ Prod.fst)) ", ") "}"
  | `(fun ($id : $ty) => $e) => paren f!"fun {paren (pp id ++ " : " ++ pp ty)} => {pp e}"
  | `(fun $id => $e) => paren f!"fun {pp id} => {pp e}"
  | `($num:numLit) => fmt (num.isNatLit?.getD 0)
  | `($str:strLit) => repr (str.isStrLit?.getD "")
  | `($fn $args*) => paren <| pp fn ++ " " ++ joinSep (args.toList.map pp) line
  | `(def $id:ident := $e) => f!"def {ppIdent id.getId} := {pp e}"
  | `(syntax $[(name := $n)]? $[(priority := $prio)]? $[$args:stx]* : $kind) => "syntax ..."  -- irrelevant for this example
  | `(macro_rules | $lhs => $rhs) => f!"macro_rules |{lhs.reprint.getD ""} => {if hideMacroRulesRhs then f!"..." else pp rhs}"
  | stx => f!"<not a core form: {stx}>"

-- integrate example expander into frontend, between parser and elaborator. Not pretty.
section Elaboration
open Lean.Elab
open Lean.Elab.Frontend

-- run expander: adapt global context and set of macro from Environment
def expanderToFrontend (ref : Syntax) (e : ExpanderM Syntax) : FrontendM Syntax := runCommandElabM <| withRef ref do
  let st ← get
  let scope := st.scopes.head!
  match e {
    gctx := fun n => (match st.env.find? n with
      | some _ => [(n, [])]
      | none   => [] : List NameRes),
    lctx := {},
    currMacroScope := st.nextMacroScope,
    macros := fun k =>
      -- our hardcoded example macros
      if k == `Lean.Parser.Term.quot then some expandStxQuot
      else if k == `Lean.Parser.Term.do then some expandDo
      else if k == `Lean.Parser.Term.paren then some expandParen
      -- `notation`, `macro`, and macros generated at runtime
      else
        let table := (macroAttribute.ext.getState st.env).table
        match table.find? k with
        | some (t::_) => some (fun stx ctx =>
          match t stx {
            mainModule := `Expander
            currMacroScope := ctx.currMacroScope
            ref := ref
            methods := Macro.mkMethods {
              expandMacro?     := fun stx =>
                try
                  let newStx ← getMacros st.env stx
                  pure (some newStx)
                catch
                  | Macro.Exception.unsupportedSyntax => pure none
                  | ex                                => throw ex
              hasDecl          := fun declName => return st.env.contains declName
              getCurrNamespace := return scope.currNamespace
              resolveNamespace? := fun n => return ResolveName.resolveNamespace? st.env scope.currNamespace scope.openDecls n
              resolveGlobalName := fun n => return ResolveName.resolveGlobalName st.env scope.currNamespace scope.openDecls n
            }
          } {
            macroScope := 0
          } with
          | EStateM.Result.ok stx s => stx
          | _ => Syntax.missing)
        | _           => none
  } (st.nextMacroScope + 1) with
  | Except.ok (stx, nextMacroScope) => do
    modify (fun st => {st with nextMacroScope := nextMacroScope})
    pure stx
  | Except.error e => do
    logError e
    pure Syntax.missing

partial def processCommand (cmd : Syntax) : FrontendM Unit := do
  let cmd' ← expanderToFrontend cmd <| expand cmd
  if cmd'.isOfKind nullKind then
    -- expander returned multiple commands => process in turn
    cmd'.getArgs.forM processCommand
  else do
    runCommandElabM <| logInfo (pp cmd')
    elabCommandAtFrontend cmd'

partial def processCommands : FrontendM Unit := do
  let cmdState    ← getCommandState
  let parserState ← getParserState
  let inputCtx    ← getInputContext
  let scope := cmdState.scopes.head!
  let pmctx := { env := cmdState.env, options := scope.opts, currNamespace := scope.currNamespace, openDecls := scope.openDecls }
  let (cmd, ps, messages) ← pure (Parser.parseCommand inputCtx pmctx parserState cmdState.messages)
  setParserState ps
  setMessages messages
  if Parser.isEOI cmd then do
    pure ()
  else do
    processCommand cmd
    processCommands

def process (input : String) (env : Environment) (opts : Options) (fileName : Option String := none) : IO (Environment × MessageLog) := do
  let fileName   := fileName.getD "<input>"
  let inputCtx   := Parser.mkInputContext input fileName
  let (_, st) ← processCommands { inputCtx := inputCtx } |>.run { commandState := Command.mkState env {} opts, parserState := {}, cmdPos := 0 }
  pure (st.commandState.env, st.commandState.messages)

def run (input : String) : CoreM Unit := do
  let env  ← getEnv
  let opts ← getOptions
  let (env, messages) ← liftM $ process input env opts
  messages.forM fun msg => do
    IO.println (← msg.toString)

end Elaboration

-- examples
-- see also `hideMacroRulesRhs` above

#eval run "
def x := 1
def e := fun (y : Nat) => x
notation \"const\" e => fun (x : Nat) => e
def y := const x
"

#eval run "
macro \"m\" n:ident : command => `(
  def f := 1
  macro \"mm\" : command => `(
    def $n:ident := f
    def f := $n:ident))
m f
mm
mm
"

end Expander
end Lean
\end{lstlisting}
\begin{lstlisting}[aboveskip=0pt,showlines=true,numbers=left,firstnumber=17]
  | ...  -- other core forms
  | _ => do
    let t ← getTransformerFor stx.getKind
    let stx ← withFreshMacroScope (t stx)
    expand stx
\end{lstlisting}
  \caption[]{Abbreviated implementation of a recursive expander for our macro system}
  \label{fig:expander}
\end{figure}
A Lean 4 implementation of the expansion algorithm described in the previous section is given in
\cref{fig:expander}; the full implementation including examples is included in the supplement.
As a generalization,
syntax transformers in the full implementation have the type \lstinline{Syntax → TransformerM Syntax} where the \lstinline{TransformerM} monad gives access to
the global context and a fresh macro scope per macro expansion. The expander itself uses an extended \lstinline{ExpanderM} monad
based on \lstinline{TransformerM} that also stores the local context and the set of registered macros.
We use the Lean equivalent of Haskell's \verb!do! notation~\cite{marlow2010haskell} to program in these monads.

As described in \cref{sec:algo}, the expander in \cref{fig:expander} has built-in
knowledge of some ``core forms'' (lines 2-16) with special expansion behavior, while all other forms are assumed to
be macros and expanded recursively (lines 19-21). Identifiers form one base case of the recursion. As described in
the previous section, they are first looked up in the local context (recall that the \lstinline{Name} of an identifier includes macro scopes),
then as a fall back in the global context plus its own top-level scopes. \lstinline{mkIdent : Name → Syntax}
creates an identifier without source information or top-level scopes, which are not needed after expansion.
\lstinline{mkOverloadedIds} implements the Lean special case of overloaded symbols to be disambiguated by
elaboration; systems without overloading support should throw an ambiguity error instead in this case.

As an example of a core binding form, the expansion of a single-parameter \lstinline{fun} is shown in lines 12-16 of \cref{fig:expander}. It
recursively expands the given parameter type, then expands the body in a new local context extended with the value of
\lstinline{id}. Here \lstinline{getIdentVal : Syntax → Name} in particular
implements the discarding of top-level scopes from binders.

Finally, in the macro case, we fetch the syntax transformer for the given node kind, run it in a new context with
a fresh current macro scope, and recurse.

\begin{figure}[ht]
\begin{lstlisting}[linerange=quoteSyntax-end,numbers=left,firstnumber=0]
import Lean

namespace Lean
namespace Expander

open Lean.Syntax

-- Result of name resolution. As in the paper, we will ignore the second component here.
abbrev NameRes := Name × List String
-- We model the global context more precisely as a mapping from symbols to qualified symbols,
-- e.g. (`a ↦ [`ns1.a, `ns2.a])
abbrev GlobalContext := Name → List NameRes

-- the simplified transformer monad
structure TransformerContext where
  gctx : GlobalContext
  currMacroScope : MacroScope

abbrev TransformerM := ReaderT TransformerContext Id
abbrev Transformer := Syntax → TransformerM Syntax

-- support syntax quotations in transformers
instance : MonadQuotation TransformerM where
  getCurrMacroScope   := do let ctx ← read; pure ctx.currMacroScope
  -- dummy impls, unused
  withFreshMacroScope := fun x => x
  getRef              := pure Syntax.missing
  withRef             := fun _ x => x
  -- The actual implementation also adds the current module name to macro scopes for global uniqueness,
  -- which we can ignore in this single-file example.
  getMainModule       := pure `Expander

-- the expander extension of the transformer monad
structure ExpanderContext extends TransformerContext where
  lctx : NameSet
  macros : Name → Option Transformer

abbrev ExpanderM := ReaderT ExpanderContext <| StateT MacroScope <| ExceptT String <| Id

instance MonadQuotation : MonadQuotation ExpanderM where
  getCurrMacroScope   := do let ctx ← read; pure ctx.currMacroScope
  withFreshMacroScope := fun x => do
    let fresh ← modifyGet (fun n => (n, n + 1))
    withReader (fun ctx => { ctx with currMacroScope := fresh }) x
  getMainModule       := pure `Expander
  -- dummy impls, unused
  getRef              := pure Syntax.missing
  withRef             := fun _ x => x

-- implicitly coerce transformer monad into expander monad
instance : Coe (TransformerM α) (ExpanderM α) where
  coe t := fun ctx => t ctx.toTransformerContext

-- simplified: ignore the module name parameter
def addMacroScope (n : Name) (scp : MacroScope) : Name :=
  Lean.addMacroScope `Expander n scp

def getGlobalContext : TransformerM GlobalContext := do
  return (← read).gctx

def getLocalContext : ExpanderM NameSet := do
  return (← read).lctx

def resolve (gctx : GlobalContext) (n : Name) : List NameRes :=
  gctx n

-- slightly more meaningful name
def getIdentVal : Syntax → Name := Syntax.getId

def getPreresolved : Syntax → List NameRes
  | Syntax.ident (preresolved := preresolved) .. => preresolved
  | _                                            => []

def mkOverloadedIds (cs : List NameRes) : Syntax :=
  Syntax.node choiceKind (cs.toArray.map (mkIdent ∘ Prod.fst))

def withLocal (l : Name) : ExpanderM Syntax → ExpanderM Syntax :=
  withReader (fun ctx => { ctx with lctx := ctx.lctx.insert l })

def getTransformerFor (k : SyntaxNodeKind) : ExpanderM (Syntax → ExpanderM Syntax) := do
  match (← read).macros k with
  | some t => return fun stx => t stx
  | none   => throw ("unknown macro " ++ toString k)

-- slightly simplified from the actual implementation
partial def getAntiquotationIds (stx : Syntax) : ExpanderM (Array Syntax) := do
  let mut ids := #[]
  for stx in stx.topDown do
    if (isAntiquot stx || isTokenAntiquot stx) && !isEscapedAntiquot stx then
      let anti := getAntiquotTerm stx
      if anti.isIdent then ids := ids.push anti
      else throw "complex antiquotation not allowed here"
  return ids

-- Get all pattern vars (as `Syntax.ident`s) in `stx`
partial def getPatternVars (stx : Syntax) : ExpanderM (Array Syntax) :=
  if stx.isQuot then
    getAntiquotationIds stx
  else match stx with
    | `(_)            => #[]
    | `($id:ident)    => #[id]
    | `($id:ident@$e) => do (← getPatternVars e).push id
    | _               => throw "unsupported pattern in syntax match"

-- expand
partial def expand : Syntax → ExpanderM Syntax
  | `($id:ident) => do
    let val : Name := getIdentVal id
    let gctx ← getGlobalContext
    let lctx ← getLocalContext
    if lctx.contains val then
      pure (mkIdent val)
    else match resolve gctx val ++ getPreresolved id with
      | []        => throw ("unknown identifier " ++ toString val)
      | [(id, _)] => pure (mkIdent id)
      | ids       => pure (mkOverloadedIds ids)
  | `(fun ($id : $ty) => $e) => do
    let val := getIdentVal id
    let ty ← expand ty
    let e ← withLocal val (expand e)
    `(fun ($(mkIdent val) : $ty) => $e)
-- end
  -- more core forms
  | `(fun $id:ident => $e) => do
    let e ← withLocal (getIdentVal id) (expand e)
    `(fun $id:ident => $e)
  | `($num:numLit) => `($num:numLit)
  | `($str:strLit) => `($str:strLit)
  | `($n:quotedName) => `($n:quotedName)
  | `($fn $args*) => do
    let fn ← expand fn
    let args ← args.mapM expand
    `($fn $args*)
  | `(def $id := $e) => do
    let e ← expand e
    `(def $id := $e)
  -- syntax: keep as-is
  | `(syntax $[(name := $n)]? $[(priority := $prio)]? $[$args:stx]* : $kind) => `(syntax $[(name := $n)]? $[(priority := $prio)]? $[$args:stx]* : $kind)
  -- macro_rules: expand rhs (but not lhs) to exercise syntax quotation macro
  | `(macro_rules | $lhs => $rhs) => do
    let vars ← getPatternVars lhs
    let rhs ← vars.foldr (fun var ex => withLocal var.getId ex) (expand rhs)
    `(macro_rules | $lhs => $rhs)
  -- we will ignore double-backtick quotations generated by `notation` for this example
  | `(``($e)) => `(`($e))
  | stx => do
    -- expansion consists of multiple commands => yield and get called back per command
    if stx.isOfKind nullKind then pure stx else do
    let t ← getTransformerFor stx.getKind
    let stx ← withFreshMacroScope (t stx)
    expand stx

open Lean.Elab.Term.Quotation
-- quoteSyntax
partial def quoteSyntax : Syntax → TransformerM Syntax
  | Syntax.ident info rawVal val preresolved => do
    let gctx ← getGlobalContext
    let preresolved := resolve gctx val ++ preresolved
    `(Syntax.ident SourceInfo.none $(quote rawVal)
        (addMacroScope $(quote val) msc) $(quote preresolved))
  | stx@(Syntax.node k args) =>
    if isAntiquot stx then pure (getAntiquotTerm stx)
    else do
      let args ← args.mapM quoteSyntax
      `(Syntax.node $(quote k) $(quote args))
  | Syntax.atom info val => `(Syntax.atom SourceInfo.none $(quote val))
  | Syntax.missing => pure Syntax.missing

def expandStxQuot (stx : Syntax) : TransformerM Syntax := do
  let stx ← quoteSyntax (stx.getArg 1)
  `(do msc ← getCurrMacroScope; pure $stx)
-- end

-- two more, simple macros
def expandDo : Transformer
  | `(do $id:ident ← $val:term; $body:term) => `(Bind.bind $val (fun $id:ident => $body))
  | _                                  => pure Syntax.missing

def expandParen : Transformer
  | `(($e)) => pure e
  | _       => pure Syntax.missing

-- custom Syntax pretty printer for our core forms that uses the paper's notation for hygienic identifiers
def ppIdent (n : Name) : Format :=
  let v := extractMacroScopes n
  fmt <| v.scopes.foldl Name.mkNum v.name

-- flip to make output more readable
def hideMacroRulesRhs := false

open Std.Format
partial def pp : Syntax → Format
  | `($id:ident) => match getPreresolved id with
    | [] => ppIdent id.getId
    | ps => ppIdent id.getId ++ bracket "{" (joinSep (ps.map (fmt ∘ Prod.fst)) ", ") "}"
  | `(fun ($id : $ty) => $e) => paren f!"fun {paren (pp id ++ " : " ++ pp ty)} => {pp e}"
  | `(fun $id => $e) => paren f!"fun {pp id} => {pp e}"
  | `($num:numLit) => fmt (num.isNatLit?.getD 0)
  | `($str:strLit) => repr (str.isStrLit?.getD "")
  | `($fn $args*) => paren <| pp fn ++ " " ++ joinSep (args.toList.map pp) line
  | `(def $id:ident := $e) => f!"def {ppIdent id.getId} := {pp e}"
  | `(syntax $[(name := $n)]? $[(priority := $prio)]? $[$args:stx]* : $kind) => "syntax ..."  -- irrelevant for this example
  | `(macro_rules | $lhs => $rhs) => f!"macro_rules |{lhs.reprint.getD ""} => {if hideMacroRulesRhs then f!"..." else pp rhs}"
  | stx => f!"<not a core form: {stx}>"

-- integrate example expander into frontend, between parser and elaborator. Not pretty.
section Elaboration
open Lean.Elab
open Lean.Elab.Frontend

-- run expander: adapt global context and set of macro from Environment
def expanderToFrontend (ref : Syntax) (e : ExpanderM Syntax) : FrontendM Syntax := runCommandElabM <| withRef ref do
  let st ← get
  let scope := st.scopes.head!
  match e {
    gctx := fun n => (match st.env.find? n with
      | some _ => [(n, [])]
      | none   => [] : List NameRes),
    lctx := {},
    currMacroScope := st.nextMacroScope,
    macros := fun k =>
      -- our hardcoded example macros
      if k == `Lean.Parser.Term.quot then some expandStxQuot
      else if k == `Lean.Parser.Term.do then some expandDo
      else if k == `Lean.Parser.Term.paren then some expandParen
      -- `notation`, `macro`, and macros generated at runtime
      else
        let table := (macroAttribute.ext.getState st.env).table
        match table.find? k with
        | some (t::_) => some (fun stx ctx =>
          match t stx {
            mainModule := `Expander
            currMacroScope := ctx.currMacroScope
            ref := ref
            methods := Macro.mkMethods {
              expandMacro?     := fun stx =>
                try
                  let newStx ← getMacros st.env stx
                  pure (some newStx)
                catch
                  | Macro.Exception.unsupportedSyntax => pure none
                  | ex                                => throw ex
              hasDecl          := fun declName => return st.env.contains declName
              getCurrNamespace := return scope.currNamespace
              resolveNamespace? := fun n => return ResolveName.resolveNamespace? st.env scope.currNamespace scope.openDecls n
              resolveGlobalName := fun n => return ResolveName.resolveGlobalName st.env scope.currNamespace scope.openDecls n
            }
          } {
            macroScope := 0
          } with
          | EStateM.Result.ok stx s => stx
          | _ => Syntax.missing)
        | _           => none
  } (st.nextMacroScope + 1) with
  | Except.ok (stx, nextMacroScope) => do
    modify (fun st => {st with nextMacroScope := nextMacroScope})
    pure stx
  | Except.error e => do
    logError e
    pure Syntax.missing

partial def processCommand (cmd : Syntax) : FrontendM Unit := do
  let cmd' ← expanderToFrontend cmd <| expand cmd
  if cmd'.isOfKind nullKind then
    -- expander returned multiple commands => process in turn
    cmd'.getArgs.forM processCommand
  else do
    runCommandElabM <| logInfo (pp cmd')
    elabCommandAtFrontend cmd'

partial def processCommands : FrontendM Unit := do
  let cmdState    ← getCommandState
  let parserState ← getParserState
  let inputCtx    ← getInputContext
  let scope := cmdState.scopes.head!
  let pmctx := { env := cmdState.env, options := scope.opts, currNamespace := scope.currNamespace, openDecls := scope.openDecls }
  let (cmd, ps, messages) ← pure (Parser.parseCommand inputCtx pmctx parserState cmdState.messages)
  setParserState ps
  setMessages messages
  if Parser.isEOI cmd then do
    pure ()
  else do
    processCommand cmd
    processCommands

def process (input : String) (env : Environment) (opts : Options) (fileName : Option String := none) : IO (Environment × MessageLog) := do
  let fileName   := fileName.getD "<input>"
  let inputCtx   := Parser.mkInputContext input fileName
  let (_, st) ← processCommands { inputCtx := inputCtx } |>.run { commandState := Command.mkState env {} opts, parserState := {}, cmdPos := 0 }
  pure (st.commandState.env, st.commandState.messages)

def run (input : String) : CoreM Unit := do
  let env  ← getEnv
  let opts ← getOptions
  let (env, messages) ← liftM $ process input env opts
  messages.forM fun msg => do
    IO.println (← msg.toString)

end Elaboration

-- examples
-- see also `hideMacroRulesRhs` above

#eval run "
def x := 1
def e := fun (y : Nat) => x
notation \"const\" e => fun (x : Nat) => e
def y := const x
"

#eval run "
macro \"m\" n:ident : command => `(
  def f := 1
  macro \"mm\" : command => `(
    def $n:ident := f
    def f := $n:ident))
m f
mm
mm
"

end Expander
end Lean
\end{lstlisting}
  \caption[]{Simplified syntax transformer for syntax quotations}
  \label{fig:quote}
\end{figure}
Syntax quotations are given as one example of a macro: they do not have built-in semantics but transform into code that
constructs the appropriate syntax tree~(\lstinline{expandStxQuot} in \cref{fig:quote}).
More specifically, a syntax quotation will, at run time (of the surrounding macro), query the current
macro scope \lstinline{msc} from the surrounding \lstinline{TransformerM} monad (code generated by \lstinline{expandStxQuot}) and apply it to all captured identifiers (code generated by \lstinline{quoteSyntax}).
\lstinline{quoteSyntax} recurses
through the quoted syntax tree, reflecting its constructors. Basic datatypes such as \lstinline{String} and
\lstinline{Name} are turned into \lstinline{Syntax} via the typeclass method \lstinline{quote}. For antiquotations,
we return their contents unreflected. In the case
of identifiers, we resolve possible global references at compile time and reflect them, while \lstinline{msc} is
applied at run time. Thus a quotation
\lstinline{`(a + $b)} inside a global context where the symbol \lstinline{a} matches declarations \lstinline{a.a} and \lstinline{b.a} is transformed to the equivalent of\\
\begin{minipage}{1.0\linewidth}
\begin{lstlisting}
do let msc ← getCurrMacroScope
   pure (Syntax.node `plus [
     Syntax.ident SourceInfo.none "a" (addMacroScope `a msc)
       [`a.a, `b.a],
     Syntax.atom SourceInfo.none "+",
     b])
\end{lstlisting}
\end{minipage}

This implementation of syntax quotations itself makes use of syntax quotations for simplicity
and thus is dependent on its own
implementation in the previous stage of the compiler. Indeed, the helper variable \lstinline{msc} must be renamed should
the name already be in scope and used inside an antiquotation.\footnote{As long as no such case exists, a hygienic implementation of syntax quotations can be bootstrapped from an unhygienic one, which is what we did in the case of Lean.} Note that \lstinline{quoteSyntax} is
allowed to reference the same \lstinline{msc} as
\lstinline{expandStxQuot} because they are part of the same macro call and the current macro scope is unchanged between them.
While alternative approaches that use fresh macro scopes on function calls \emph{within} a macro are thinkable, we prefer the presented behavior, which matches that of the Scheme family, because it preserves referential transparency: if \lstinline{quoteSyntax} is inlined into \lstinline{expandStxQuot}, the behavior is unchanged.

\subsection{Extended Quasiquotations}
\label{sec:quot}

Automatic hygiene can greatly simplify development of macros, but a convenient way for con- and destructing syntax is at least as important.
Before we get to more complex macro examples below, we want to describe some syntactic extensions to quotations and antiquotations we have
implemented that will come in useful.

A first obvious such extension is to allow quotations including antiquotations as patterns such as after \lstinline{match} or \lstinline{fun}.
\begin{lstlisting}
fun
  | `(())       => ...
  | `(($e))     => ...
  | `(($e, $f)) => ...
\end{lstlisting}
Because every use of patterns eventually unfolds to a \lstinline{match} in Lean, this is in fact implemented as a macro that expands \lstinline{match} terms with quotation patterns into ones without such patterns.
Note also that because no new syntactic identifiers are generated while matching against a quotation, there is no issue of hygiene in this case.

For syntax with repeated parts, quotation \emph{splices} enable us to match or introduce these parts as a whole.
For example, a recursive macro for $n$-tuple syntax can be written as
\begin{lstlisting}
macro_rules
  | `(())          => `(Unit.unit)
  | `(($e))        => e
  | `(($e, $es,*)) => `(Prod.mk $e ($es,*))
\end{lstlisting}
Here \lstinline{$es,*} matches/introduces the remaining elements of the tuple, including its separators.
Analogous splicing syntax exists for other separators, as well as \lstinline{$x*} for separator-less iteration.
At most one splice can be used per sequence, but it can be pre- and suffixed with an arbitrary (but fixed) number of other elements.

In \lstinline{$x*}, \lstinline{x} has type \lstinline{Array Syntax}, the same type as the second argument of the \lstinline{node} constructor.
For \lstinline{$x,*} and similar we instead use the dependent wrapper type \lstinline{SepArray ","} that provides convenience access functions for the sequence both with and without separator elements.
Finally, we also provide implicit coercions between these types that automatically insert/remove/replace the separators accordingly.

While exposing splices as typed values in this way ensures that we can comfortably process or synthesize them procedurally as well, it is often more convenient to inspect splice contents immediately as part of the quotation.
For this we support extended splices \lstinline{$[...]*} etc.\ where the splice content is parsed like an element of the sequence and can contain nested antiquotations.
If used as a pattern, the match succeeds if and only if the nested pattern matches every element, in which case the contained antiquotations are each bound to an \lstinline{Array} of all corresponding element-wise matches.
\begin{lstlisting}
match stx with
| `(match $discr with $[| $patss,* => $branches]*) =>
  -- discr : Syntax
  -- patss : Array (SepArray ",")
  -- branches : Array Syntax
  ...
\end{lstlisting}

By default, quotations are parsed as either terms or top-level commands, since these syntactic categories are both commonly used and should usually be disjoint.
Other syntactic categories, e.g.\ the category of universe \lstinline{level}s that heavily overlaps with \lstinline{term}, can be specified explicitly at the beginning of a quotation.
Similarly, antiquotations can be suffixed with a colon followed by a category or named parser where otherwise ambiguous.
\begin{lstlisting}
match levelStx with
| `(level| $id:ident) => ...  -- a universe variable
| `(level| _)         => ...  -- a universe placeholder
| `(level| $l)        => ...  -- any (other) universe term
\end{lstlisting}

For a full example of using these and other features, we can look at a macro rule unfolding syntax such as
\begin{lstlisting}
fun
  | some a, some b => some (a + b)
  | _,      _      => none
\end{lstlisting}
into\\
\begin{minipage}{\linewidth}
\begin{lstlisting}
fun x.1 x.2 =>
  match x.1, x.2 with
  | some a, some b => some (a + b)
  | _,      _      => none
\end{lstlisting}
\end{minipage}

\begin{figure}[tb]
\begin{lstlisting}[aboveskip=0pt,showlines=true,numbers=left]
macro_rules
  | `(fun
        | $ps1,* => $rhs1
        $alts:matchAlt*) => do
    let discrs ← ps1.getElems.mapM (fun _ => withFreshMacroScope `(x))
    `(fun $discrs* =>
        match $[$discrs:ident],* with
        | $ps1,* => $rhs1
        $alts:matchAlt*)
\end{lstlisting}
  \caption[]{A macro rule for expanding a combined fun-match syntax.}
  \label{fig:fun-match}
\end{figure}
The macro rule (\cref{fig:fun-match}) derives the number of \emph{discriminants} (\lstinline{x.1, x.2} in the example) from the number of patterns of the first alternative; if other alternatives have differing number of patterns, it will lead to an elaboration error in \lstinline{match} later on. The macro then introduces a lambda abstraction over a sequence of fresh variable names of this number and subsequently matches on them using the given patterns.

It does so by matching on the first alternative of the match in detail, then capturing the remaining ones in \lstinline{alts : Array Syntax}.
The left-hand side \lstinline{ps1} of the first alternative is a \lstinline{SepArray ","}, so we use \lstinline{getElems} to access its elements and generate a fresh variable for each of them, which we do by running the single quotation \lstinline{`(x)} repeatedly under \lstinline{withFreshMacroScope}, annotating the variable with a unique macro scope each time\footnote{This is comparable to a call to the \lstinline{gensym} function found in many Lisp systems.}.

With \lstinline{discrs : Array Syntax} generated, we insert it into the final quotation, once as a straight sequence after \lstinline{fun} and once separated by commas after \lstinline{match}, followed by the alternatives copied from the input without changes.
Note that \lstinline{$discrs:ident*} would not have worked in this case because identifiers are merely a special case of the more general \lstinline{match} discriminant syntax that allows prefixing a discriminant with \lstinline{h:}, where the identifier \lstinline{h} will then hold the proof that the discriminant matched the corresponding pattern.
Thus there is no direct identifier sequence to insert and we have to instead say that we insert a sequence of general discriminants, each one built up of an identifier without a proof variable prefix, which internally will wrap each element in an additional syntax tree node of the \lstinline{matchDiscr} kind.
This necessary disambiguation of overlapping syntax sadly is a price we have to pay for our preference of such syntax over more regular but verbose one such as S-expressions.

Finally, for the sake of completeness we will mention the \emph{token antiquotation} \lstinline{
This kind of antiquotation is mostly useful for displaying errors on specific tokens and preserving metadata in transformations. \\
\begin{minipage}{\linewidth}
\begin{lstlisting}
| `(tactic| case $tag =>%$arrTk $tac) => do
  ...
  reportUnsolvedGoalsAt arrTk

...

  case cons => skip
          --^ unsolved goals displayed here
\end{lstlisting}
\end{minipage}

\section{Integrating Macros into Elaboration}
\label{sec:elab}

The macro system as described so far can handle most syntax sugars of Lean 3 except for ones requiring type information.
For example, the \emph{anonymous constructor} \lstinline{⟨e, ...⟩} is sugar for \lstinline{(c e ...)} if the expected type
of the expression is known and it is an inductive type with a single constructor \lstinline{c}. While trivial to parse,
there is no way to implement this syntax as a macro if expansion is done strictly prior to elaboration. A more complex
example is the \emph{structure instance notation} \lstinline!{ field1 := e, ...}! that must analyze the definition of the
given or inferred structure type in order to expand to the correct constructor call. To the best of
our knowledge, none of the ITPs listed in the introduction support hygienic elaboration extensions of this kind, but we
will show how to extend their common elaboration scheme in that way in this section.

Elaboration\footnote{At the term level; elaboration of other syntactic categories work analogously but with different output types.} can be thought of
as a function \lstinline{elabTerm : Syntax → ElabM Expr} in an appropriate monad \lstinline{ElabM}\footnote{Or some
  other encoding of effects.} from a (concrete or abstract) surface-level syntax tree type \lstinline{Syntax} to a fully-specified core
term type \lstinline{Expr}~\cite{moura2015elaboration}. We have presented the (concrete) definition of \lstinline{Syntax} in Lean
4 in \cref{sec:impl}; the particular definition of \lstinline{Expr} is not important here.
While such an elaboration system could readily be composed with a type-insensitive macro expander such as the one presented in \cref{sec:hygge}, we would
rather like to \emph{intertwine} the two to support type-sensitive but still hygienic-by-default macros (henceforth
called \emph{elaborators}) without having to reimplement macros of the kind discussed so far. Indeed, these can automatically
be adapted to the new type given an adapter between the two monads, similarly to the adaption of macros to
\emph{expanders} in \cite{dybvig1986expansion}:
\begin{lstlisting}
def transformerToElaborator (t : Syntax → TransformerM Syntax) :
    Syntax → ElabM Expr :=
  fun stx => do
    let stx' ← (transformerMToElabM t) stx
    elabTerm stx'
\end{lstlisting}
Because most parts of our hygiene system are implemented by the expander for syntax quotations, the only changes to an
elaboration system necessary for supporting hygiene are storing the current macro scope in the elaboration monad (to be passed
to the expansion monad in the adapter) and allocating a fresh macro scope whenever a macro or elaborator is invoked. Thus elaborators immediately benefit from hygiene
as well whenever they use syntax quotations to construct unelaborated helper syntax objects to pass to \verb!elabTerm!.
In order to support syntax quotations in these two and other monads, we generalize their implementation to a new monad
typeclass implemented by both monads.
\begin{lstlisting}
class MonadQuotation (m : Type → Type) where
  getCurrMacroScope : m MacroScope
  withFreshMacroScope : m α → m α
\end{lstlisting}
The second operation is not used by syntax quotations directly, but can be used by procedural macros and elaborators to manually enter
new macro call scopes.

As an example, the following is a simplified implementation of the anonymous constructor syntax mentioned above.
\begin{lstlisting}
@[termElab anonymousCtor] def elabAnonymousCtor : Syntax → ElabM Expr
  | `(⟨$args,*⟩) => do
    let expectedType ← getExpectedType
    match Expr.getAppFn expectedType with
    | Expr.const constName _ _ => do
      let ctors ← getCtors constName
      match ctors with
      | [ctor] => do
        let stx ← `($(mkCIdent ctor) $args*)
        elabTerm stx
  ...  -- error handling
\end{lstlisting}
The \lstinline{[termElab]} attribute registers this elaborator for the given syntax node kind.
The function \lstinline{mkCIdent : Name → Syntax}
synthesizes a hygienic reference to the given constant name by storing it as a top-level scope and applying a reserved
macro scope to the constructed identifier.
Note the monadic binding of the syntax quotation, and that the separators of \lstinline{$args,*} are implicitly discarded when it is used as a plain sequence \lstinline{$args*}.

This implementation fails if the
expected type is not yet sufficiently known at this point. The actual implementation\footnote{\url{https://github.com/leanprover/lean4/blob/IJCAR20-LMCS/src/Lean/Elab/BuiltinNotation.lean\#L16}} of this elaborator extends the code
by \emph{postponing} elaboration in this case. When an elaborator requests postponement,
the system returns a fresh metavariable as a placeholder and associates the input syntax tree with it. Before
finishing elaboration of a command, postponed elaborators associated with unsolved metavariables are retried until they all
ultimately succeed, or else elaboration is stuck because of cyclic dependencies and an error is signed.

\section{Tactic Hygiene}
\label{sec:tactic}

Lean 3 includes a tactic framework that, much like macros, allows users to write custom automation either procedurally
inside a tactic monad or ``by example'' using tactic language quotations, or in a mix of both~\cite{tactic}.
For example, Lean 3 uses a short tactic block to prove injection lemmas for data constructors.
\begin{lstlisting}
def mkInjEq : TacticM Unit :=
`[intros; apply propext; apply Iff.intro; ...]
\end{lstlisting}
Unfortunately, this code unexpectedly broke in Lean 3 when used from a library for homotopy type theory that defined its own
\lstinline{propext} and \lstinline{Iff.intro} declarations;\footnote{\url{https://github.com/leanprover/lean/pull/1913}} in other
words, Lean 3 tactic quotations are unhygienic and required manual intervention in this case. Just like with
macros, the issue with tactics is that binding structure in such embedded terms is not known at declaration time. Only
at tactic run time do we know all local variables in the current context that preceding tactics may have added or
removed, and therefore the scope of each captured identifier.

Arguably, the Lean 3 implementation also exhibited a lack of hygiene in the handling of tactic-introduced identifiers:
it did not prevent users from referencing such an identifier outside of the scope it was declared in.
\begin{lstlisting}
def myTac : TacticM Unit := `[intro h]
lemma triv (p : Prop) : p → p := begin myTac; exact h end
\end{lstlisting}

Coq's similar Ltac tactic language~\cite{delahaye:00} exhibits the same issue and users are advised not to introduce
fixed names in tactic scripts but to generate fresh names using the \lstinline{fresh} tactic
first,\footnote{\url{https://github.com/coq/coq/issues/9474}} which can be considered a manual hygiene solution.

Lean 4 instead extends its automatically hygienic macro implementation to tactic scripts by allowing regular macros in the place of
tactic invocations.
\lstinputlisting[linerange=myTac-end]{supplement/Examples.lean}
By the same hygiene mechanism described above, introduced identifiers such as \lstinline{h} are renamed so as not to be
accessible outside of their original scope, while references to global declarations are preserved as top-level scope
annotations. Thus Lean 4's tactic framework resolves both hygiene issues discussed here without requiring manual
intervention by the user. Expansion of tactic macros in fact does not precede but is integrated into the \emph{tactic
  evaluator} \lstinline{evalTactic : Syntax → TacticM Unit} such that recursive macro calls are expanded lazily,
allowing for combinators like \lstinline{repeat} that would otherwise lead to infinite recursion during expansion.
\begin{lstlisting}
syntax "repeat" tactic : tactic
macro_rules
  | `(tactic| repeat $t) => `(tactic| try ($t; repeat $t))
\end{lstlisting}
Note that \lstinline{macro} cannot be used here because the parser for \lstinline{repeat} would not yet be available
in the right-hand side. When \lstinline{$t} eventually fails, the recursion is broken without visiting and expanding the
subsequent \lstinline{repeat} macro call. The \lstinline{try} tactical is used to ignore this eventual failure.

While we believe that macros will cover most use cases of Lean 3's tactic quotations in Lean 4, their use
within larger \lstinline{TacticM} metaprograms can be recovered by passing such a quotation to \lstinline{evalTactic}:
\lstinputlisting[linerange=myTac2-end]{supplement/Examples.lean}
\lstinline{TacticM} implements the \lstinline{MonadQuotation} typeclass for this purpose.

\section{Best-Effort Eager Name Analysis in Macros}
\label{sec:eager}

The dynamic nature of binding in macros has enabled us to implement many Lean language features as macros, with hygiene guaranteeing that bindings within the macro do not interfere with ones outside of it.
However, while knowing that a mistyped identifier in a macro will not be accidentally be bound by bindings at the use site is great in theory, in practice it would be even better to be told about the typo immediately while writing the macro!
This is especially true when renaming a declaration, either manually where we might accidentally miss an occurrence of it inside a macro and then must track name binding errors at use sites back to the responsible macro, or automatically where refactoring tools must treat macros as black boxes.
After all, a static view of a notation's binding structure is exactly what we gave up in \cref{sec:intro} in exchange for the ability to arbitrarily abstract over bindings.
For example, there is no general way to statically analyze whether \lstinline{x} inside the following quotation is well-scoped given an arbitrary computation resulting in \lstinline{stx}:
\begin{lstlisting}
let stx ← ...
`(fun $stx => x)
\end{lstlisting}
With this theoretical limitation in mind, and given that this is more of a practical issue of maintenance, perhaps a practical, best-effort solution is sufficient as long as we retain all hygiene guarantees.
We have done so with an \emph{opt-in}, partial but extensible approach to eager name resolution in macros based on a variant of our quotation syntax.
\begin{lstlisting}
``(fun x => x + $y + id z)  -- error: unknown identifier 'z'
\end{lstlisting}
A \emph{double-backtick} quotation eventually unfolds to the basic, single-backtick version and thus retains all its semantics.
Before that, however, it recursively checks for identifiers that can statically be assumed to be unbound using the following heuristics:
\begin{enumerate}
  \item If there is a special \emph{precheck} hook registered for the syntax kind in question, we use it.
        Precheck hooks can signal binding errors as well as recursively continue the precheck on nested syntax, possibly with an extended (untyped) \emph{quotation context}, which is initially empty.
        For example, we provide a precheck hook for \lstinline{fun x => e} that recurses into \lstinline{e} after adding \lstinline{x} to the quotation context.
        The central \emph{identifier} precheck hook ultimately raises an error if an identifier is reached that is neither in the global, extra-macro context, nor in the quotation context.
        Other examples for precheck hooks we have added are for \lstinline{match} and function application.
  \item Otherwise, if there are no identifiers in the quoted syntax, we assume that there is no risk of unbound ones, and the check succeeds.
        In particular, antiquotations (which contain \emph{unquoted} identifiers only) are always skipped.
  \item Otherwise, if the quoted syntax is a macro, we unfold it and precheck the result.
        Here we assume that the macro is sufficiently good-natured: while macros are pure functions by definition in Lean, their behavior and in particular binding structure could in theory change drastically enough in the presence of antiquotations that it could lead to false positives of this analysis.
        If that is the case, the user either has to provide a custom precheck hook for it, or fall back to the basic quotation syntax.

        In contrast to macros, we cannot typically run elaborators directly during this check because they usually depend on type information that is not yet available, and are not guaranteed to be pure.
  \item Otherwise the check fails since we do not know how to analyze the syntax at hand.
        Again, users would either have to provide a precheck hook, or use to the single-backtick quotation syntax.
\end{enumerate}
This approach is clearly best-effort with many opportunities for unhandled cases.
However, since it is guaranteed that after the check the semantics are the same as for the basic quotation syntax, including unchanged hygiene guarantees, we believe that the practical advantages of using the syntax where possible are significant.
In particular, we have observed that quotations capturing global identifiers, which are most at risk of breaking during refactorings, are usually quite simple while complicated quotations that are used to translate one syntax into a more general one often do not reference any identifiers at all, and thus would not benefit from the additional check anyway.

This is especially true for notations, which are usually exceedingly simple in structure, most often consisting of nothing but an application of a global function symbol to the notation arguments, which is covered by the identifier and application precheck hooks.
Thus we have modified the \lstinline{notation} macro to use double-backtick quotations by default in order to make users immediately aware of any unbound identifiers inside of them, restoring its Lean 3 behavior (where the absence of macros naturally allowed for such a check).
\begin{lstlisting}
notation "∃" x "," e => Exits.intro (fun x => e)  -- error: unknown identifier 'Exits.intro'
\end{lstlisting}
We provide an option to disable this change, though all notations in the standard library passed the additional check without modification.
The check did on the other hand find a notation example in our documentation that contained an accidentally unbound identifier.

\section{Related Work}
\label{sec:related}

The main inspiration behind our hygiene implementation was Racket's \emph{Sets of Scopes}~\cite{sets-of-scopes}
hygiene algorithm. Much like in our approach, Racket annotates identifiers both with scopes from their original context
as well as with additional macro scopes when introduced by a macro expansion. However, there are some significant
differences: Racket stores both types of scopes in a homogeneous, unordered set and does name resolution via a maximum-subset
  check. For both simplicity of implementation and performance, we have reduced scopes to the bare minimal
  representation using only strict equality checks, which we can easily encode in our existing \lstinline{Name}
  implementation. In particular, we only apply scopes to matching identifiers and only inside syntax quotations. This
  optimization is of special importance because top-level declarations in Lean and other ITPs are not part of a single,
  mutually recursive scope as in Racket, but they each open their own scope over all subsequent declarations, which would
  lead to a total number of scope annotations quadratic in the number of declarations using the Sets of Scopes algorithm.
  Finally, Racket detects macro-introduced identifiers using a ``black-box'' approach without the
  macro's cooperation following the marking approach of~\citet{kohlbecker1986hygienic}: a fresh macro scope is applied to all
  identifiers in the macro input, then inverted on the macro output.
  While elegant, a naive implementation of this approach can again result in
  quadratic runtime compared to unhygienic expansion and requires further optimizations in the form of lazy scope
  propagation \citet{dybvig-hyg}, which is difficult to implement in a pure language like Lean.
  Our ``white-box'' approach based on the single primitive of an effectful syntax quotation, while slightly easier to escape
  from in procedural syntax transformers, is simple to implement, incurs minimal overhead, and is equivalent for pattern-based
  macros.

The idea of automatically handling hygiene in the macro, and not in the expander, was introduced in~\cite{clinger1991macros}, though only for
pattern-based macros. MetaML~\cite{taha2000metaml} refined this idea by tying hygiene more specifically to syntax
quotations that could be used in larger metaprogram contexts, which Template Haskell~\cite{sheard2002template}
interpreted as effectful (monadic) computations requiring access to a fresh-names generator, much like in our design.
However, both of the latter systems should perhaps be characterized more as metaprogramming frameworks than Scheme-like macro
systems: there are no ``macro calls'' but only explicit splices and so only built-in syntax with known binding semantics
can be captured inside syntax quotations. Thus the question of which captured identifiers to rename becomes trivial
again, just like in the basic notation systems discussed in \cref{sec:intro}.

While the vast majority of research on hygienic macro systems has focused on S-expression-based languages, there have
been previous efforts on marrying that research with non-parenthetical syntax, with different solutions for combining syntax tree
construction and macro expansion. The Dylan language requires macro syntax to use predefined terminators
and eagerly scans for the end of a macro call using this knowledge~\cite{dylan}, while in Honu~\cite{honu} the syntactic structure of
a macro call is discovered during expansion by a process called ``enforestation''. The Fortress~\cite{fortress} language
strictly separates the two concerns into grammar extensions 
and transformer declarations, much like we do. Dylan and Fortress are restricted to pattern-based macro declarations and
thus can make use of simple hygiene algorithms while Honu uses the full generality of the Racket macro expander. On the
other hand, Honu's authors ``explicitly trade expressiveness for syntactic simplicity''~\cite{honu}. In order to express the full
Lean language and desirable extensions in a macro system, we require both unrestricted syntax of macros and procedural
transformers.

Many of the antiquotation extensions presented in \cref{sec:quot} have been inspired by similar syntax in Rust's pattern-based macro language, though we have opened their use to procedural macros as well, using appropriate type representations.

Many theorem provers such as Agda, Coq, Idris, and Isabelle not already based on
a macro-powered language provide restricted syntax extension mechanisms,
circumventing hygiene issues by statically determining binding as seen in \cref{sec:intro}. Extensions
that go beyond that do not come with automatic hygiene guarantees.
Agda's macros,\footnote{\url{https://agda.readthedocs.io/en/v2.6.0.1/language/reflection.html\#macros}} for example,
operate on the De Bruijn index-based core term level and are not
hygienic.\footnote{\url{https://github.com/agda/agda/issues/3819}}
The ACL2 prover in contrast uses a subset of Common Lisp as its input language and adapts the hygiene algorithm of
\citet{dybvig-hyg} based on renaming~\cite{acl2}. The experimental Cur~\cite{chang2019dependent} theorem prover is a
kind of dual to our approach: it takes an established language with hygienic macros, Racket, and extends it with a
dependent type system and theorem proving tools. ACL2 does not support tactic scripts, while in Cur they can be defined
via regular macros. However, this approach does not currently provide tactic hygiene as defined in
\cref{sec:tactic}.\footnote{\url{https://github.com/wilbowma/cur/issues/104}}
The Eisbach~\cite{matichuk2016eisbach} proof method language for Isabelle is notable in that while it allows for
reusable proof methods (comparable to Lean's tactics) to be abstracted over terms, these terms are analyzed and typechecked
before being passed to the proof method, and there is no interaction between names in different proof methods either
when using the ``structured'' proof style, so the question of hygiene is moot.
Using the ``unstructured'' proof style, the same hygiene issue as noted for other system can be triggered.
\begin{lstlisting}[morekeywords={method,apply}]
method my_allI =
  rule_tac allI, rename_tac escaped

lemma "∀ x. x = x"
  apply my_allI
  apply (rule_tac t = escaped in refl)
\end{lstlisting}



\section{Conclusion}

We have proposed a new macro system for interactive theorem provers that enables syntactic abstraction and reuse far
beyond the usual support of mixfix notations. Our system is based on a novel hygiene algorithm designed with a focus on
minimal runtime overhead as well as ease of integration into pre-existing codebases, including integration into standard
elaboration designs to support type-directed macro expansion. Despite that, the algorithm is general enough to provide
a complete hygiene solution for pattern-based macros and provides flexible hygiene for procedural macros. We have also demonstrated
how our macro system can address unexpected name capture issues that haunt existing tactic frameworks. We have
implemented our method in the new version of the Lean theorem prover, Lean 4; it should be sufficiently attractive and
straightforward to implement to be adopted by other interactive theorem proving systems as well.

\paragraph{Acknowledgments.}
  We are very grateful to
  the anonymous reviewers,
  David Thrane Christiansen,
  Gabriel Ebner,
  Matthew Flatt,
  Sebastian Graf,
  Alexis King,
  Daniel Selsam,
  and
  Max Wagner
  for extensive comments, corrections, and advice.

\bibliographystyle{alphaurl}
\bibliography{macro}

\end{document}